\documentclass{jfm}
\usepackage{graphicx}
\usepackage{epstopdf, epsfig}
\usepackage{amsmath}
\usepackage{adjustbox}
\usepackage{upgreek}
\usepackage{comment}
\usepackage{booktabs}
\usepackage{amssymb}
\usepackage{makecell}
\usepackage{color}
\usepackage{rotating}
\usepackage{pdflscape}

\usepackage{hyperref}
\hypersetup{
    colorlinks=true,
    linkcolor=black,
    citecolor=black,
    urlcolor=black
}
\usepackage{tikz}
\usetikzlibrary{spy}
\usepackage{dashrule}

\usepackage{cancel}
\usepackage{xcolor}

\newcommand{\UMEAN}{\overline{U}}

\newcommand{\TAUIJ}{\tau^{\mathrm{\text{SGS}}}_{ij}}

\newcommand{\thetaref}{\langle \vartheta \rangle}
\newcommand{\Vin}{$V_{\textrm{in}}\,\,$}

\shorttitle{Asymmetric turbulence behaviour in wind farm wakes caused by the Coriolis force}
\shortauthor{G. Centurelli, J. Peinke, B. Djath, J. Schulz-Stellenfleth and G. Steinfeld}

\title{On the asymmetric behaviour of turbulence in the wake of large offshore wind farms caused by the Coriolis force}

\author{Gabriele Centurelli\aff{1,2} \corresp{\email{gabriele.centurelli@uni-oldenburg.de}}
    , Joachim Peinke\aff{1}
    , Bughsin' Djath \aff{3}
    , Johannes Schulz-Stellenfleth\aff{3}  
    \and Gerald Steinfeld\aff{1,2} }

\affiliation{\aff{1}Carl von Ossietzky Universit\"at Oldenburg, School of Mathematics and Science, Institute of Physics
\aff{2}ForWind - Center for Wind Energy Research, K\"upkersweg 70, 26129 Oldenburg, Germany 
\aff{3} Helmholtz-Zentrum Hereon, Institute of Coastal Systems - Analysis and Modeling, 21502 Geesthacht, Germany}

\begin{document}
\maketitle

\begin{abstract}
Large offshore wind farm wakes in shallow atmospheric boundary layers (ABL) exhibit often an asymmetric behaviour when observed through Synthetic-Aperture-Radar or simulated through Large-Eddy Simulations (LES).
In previous LES of wind farms in the northern hemisphere, the asymmetry manifests as a streak at the left side of the wake, looking downstream, where the turbulence kinetic energy (TKE) is greater than the surrounding flow.
This work aims at clarifying the physical mechanism that leads to the formation of such a phenomenon.
Identifying the Coriolis force as one possible source of asymmetry in the resolved physics, we simulate a real wind farm located in the German Bight operating under different ABLs: one representative of the northern hemisphere; one of the southern hemisphere; and three fictitious ABLs where the Coriolis effects on the inflow and wake, i.e. veer and the wake deflecting force, are removed individually or altogether.
Our results show that the TKE streak appears on the opposite side of the wake, i.e. the right one, in the southern hemisphere, and it is primarily caused by veer in the incoming flow, a result of the Coriolis force in a marine ABL. The process involves a larger TKE production which originates from a larger vertical shear promoted where the undisturbed veer profile converges towards the wake in the top part of the ABL.
We find that the TKE streak improves the farm wake recovery modestly.
Finally, we compare the asymmetry modelled by LES with those observed in several on-field measurements, finding striking similarities.

\end{abstract}



\section{Introduction}
\label{sec:intro}
To reach carbon neutrality by 2050, the current European wind energy capacity installed offshore will increase by multiple folds over its current value \citep{europeancommission2019greendeal}. The higher wind speeds and the more consistent conditions at sea allow for a higher capacity factor at the offshore wind farms compared to those installed on land \citep{Archer2005}. However, the costs associated with cabling offshore power plants and the very limited sea surface area that is not already claimed by fishing, marine traffic, or military activities, promoted a development based on clusters \citep{iea2019offshorewind}. Therefore, several hundreds of wind turbines are all cabled to a central hub directly connected to the grid. 

The huge costs for venturing further and further away from the coast \citep{Voormolen2016} and the relatively limited area where a favourable wind resource meets sea bed depths suitable to either floating or monopile wind turbine installation \citep{Messmer2023} lead to several wind farm clusters operating at a close distance from each other. This situation is unfavourable because, as shown by \cite{Platis.2018}, the wind farm clusters modify the atmospheric boundary layer by injecting long-lasting wakes. In numerous observations, 70\,km was not a sufficient downstream distance to allow the full recovery of a cluster wake. Although a significant statistic has not been established yet, the flight measurements presented in \citet{zumBerge2024} and \cite{Canadillas.2020} show that under stably stratified atmospheric boundary layers (ABL), the current wind farm clusters installed in the German Bight can cause velocity deficits with respect to the free stream in the order of magnitude of 20\% and 10\% at distances of 30 and 40\,km downstream in their wake, respectively. This implies that wake interactions already happen across the wind farm clusters installed (e.g.) in the German Bight, and their impact on the overall yield of the area will be larger as the expansion of offshore wind energy is realised.\\
Understanding cluster wakes is essential not only to reach a higher ratio between the actual energy yields of wind farm clusters and the installed capacity, but also to understand the impact on the climate in the areas hosting the expansion.

\begin{figure}
a) \begin{tikzpicture}[spy using outlines={rectangle,black,magnification=3, connect spies}]
\node {\pgfimage[interpolate=true, width=0.46\textwidth]{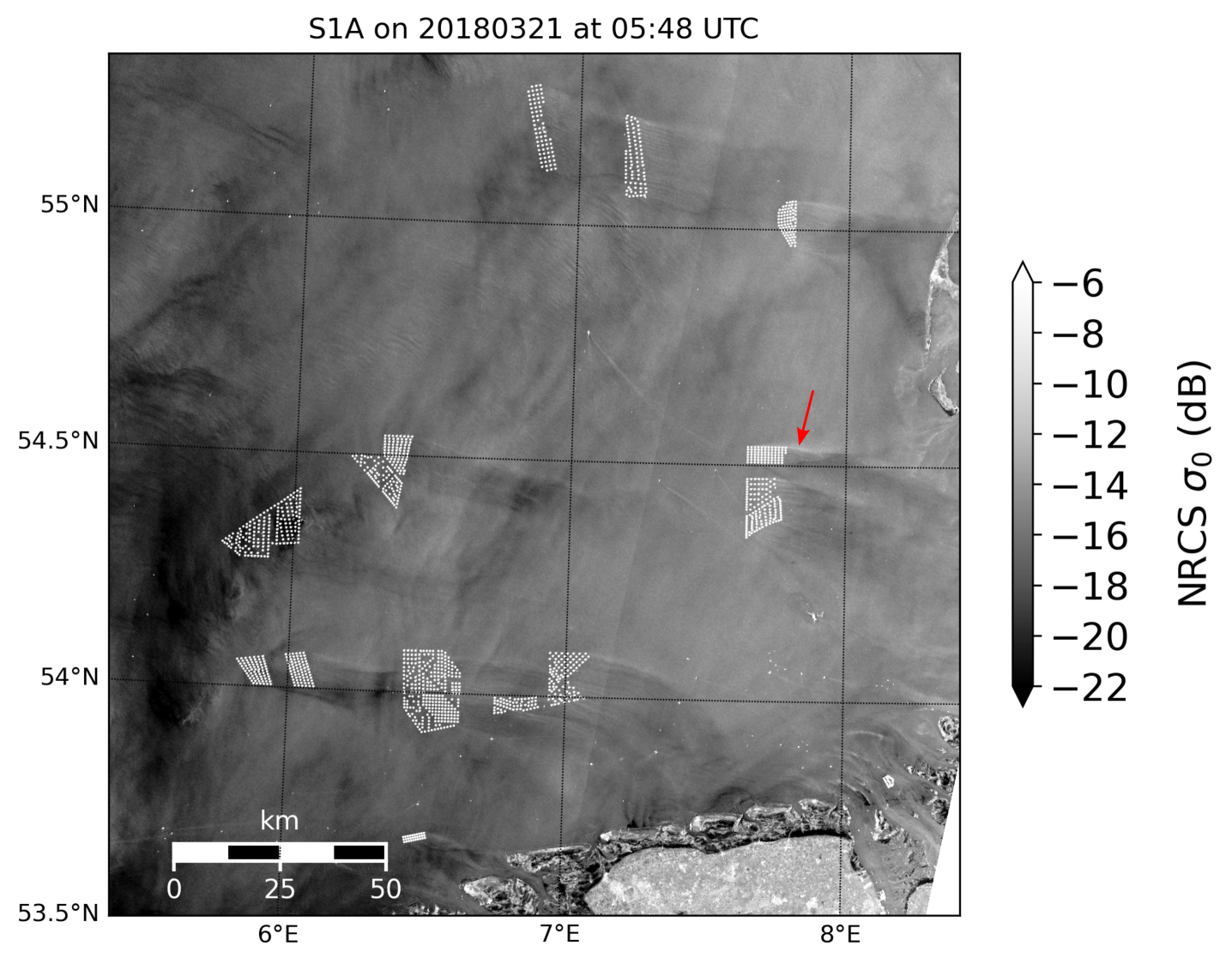}};
\spy[rectangle, black, width=2cm, height=1cm] on (0.9,0.15) in node [left] at (-0.3,0.9);
\end{tikzpicture}%
b) \includegraphics[scale=.35]{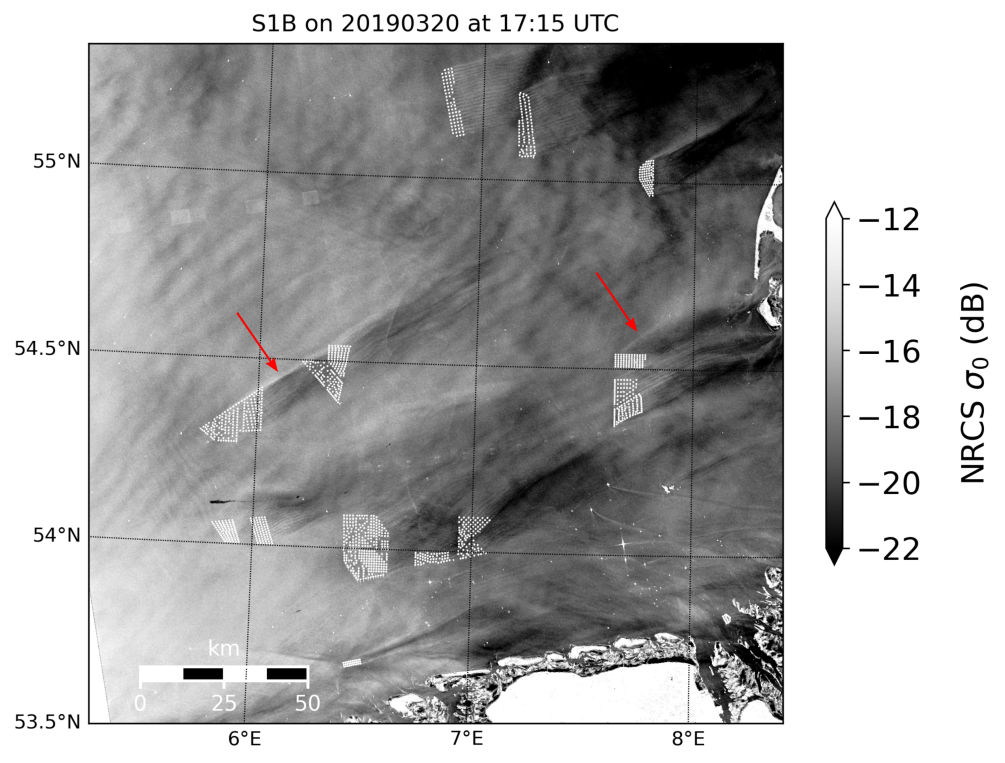}\\
c) \includegraphics[scale=.35]{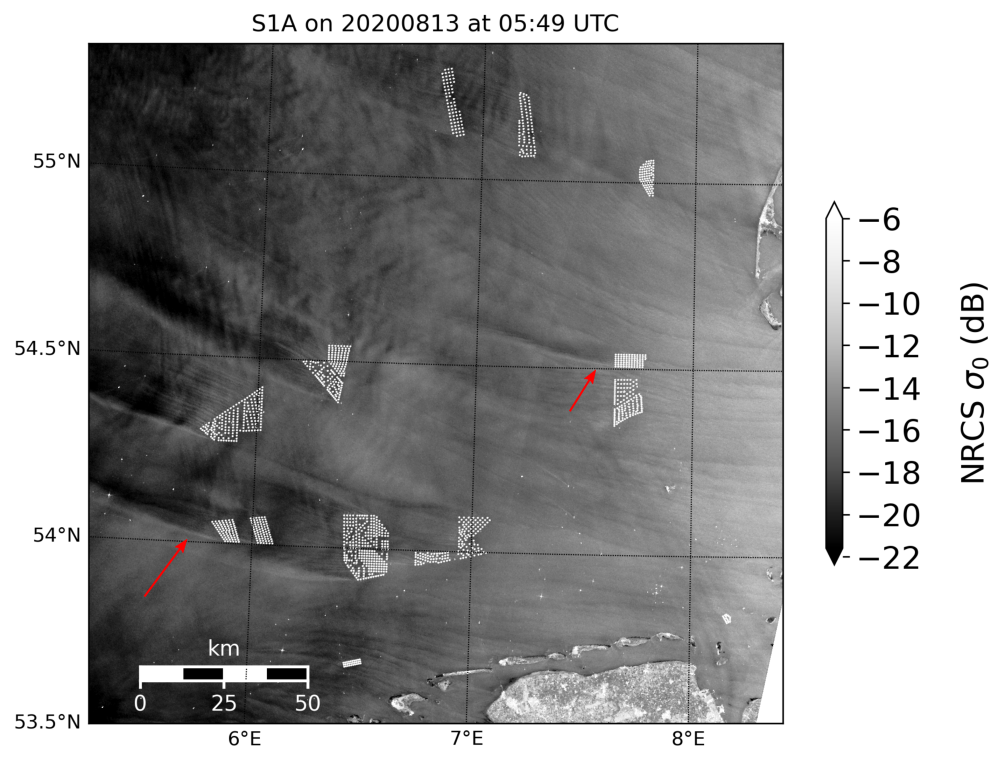}
d) \includegraphics[scale=.35]{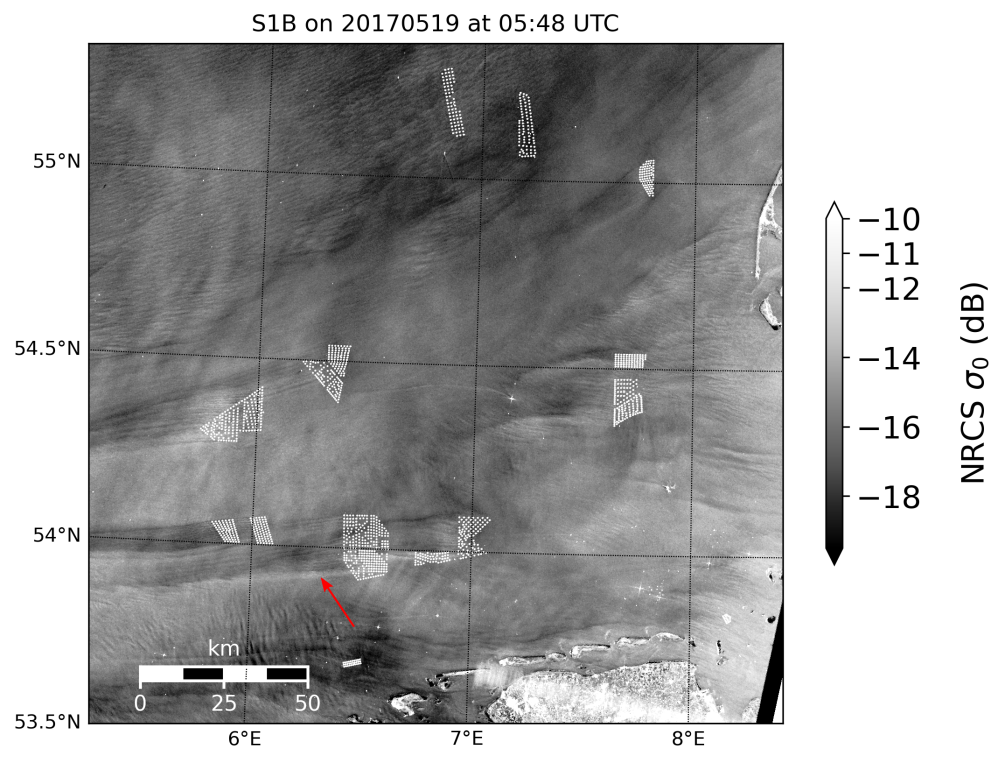}
\caption{Normalized Radar Cross Section (NRCS) for several situations captured by Copernicus satellite Sentinel-1A/B when passing over the German Bight, acquired on 21 March 2018 at 05:48 UTC (a), 20 March 2019 at 17:15 UTC (b), 13 August 2020 at 05:49 UTC (c), and 19 May 2017 at 05:48 UTC (d). In cases of westerly winds, the wakes generated by the wind farms display a streak of enhanced NRCS along the northern wake boundary (panels a,b). On the other hand, under easterly winds, a similar streak occurs along the southern boundary (panels c,d). Red arrows point at the most striking feature in each scene. The magnifying window in panel (a) highlights the wind farm and situation we aimed at replicating in the numerical set-up.}
\label{fig:SAR_panels}
\end{figure}

A phenomenon that remains poorly understood is the observed asymmetric behaviour of the cluster wake 
in several images collected by Synthetic-Aperture-Radar (SAR). This remote sensing technique, better detailed in \S\ref{sec:SAR}, measures the radar backscatter at the sea surface as Normalized Radar Cross Section (NRCS). The NRCS is affected by the sea surface capillary waves, thus it directly correlates to the friction velocity $u^*$ \citep{Durden1985}. \\
The four panels of Fig. \ref{fig:SAR_panels}, displaying the NRCS fields sampled by the \emph{Copernicus Sentinel 1} A/B satellites when passing over the German Bight, show a lower NRCS in the wake of the clusters. While a streak of larger NRCS can be observed at the northern edge of the wakes of the offshore wind farms \emph{Amrumbank West}, \emph{Sandbank}, and \emph{DanTysk} in cases with westerly winds (Fig. \ref{fig:SAR_panels} (a)), and \emph{N6} and \emph{N8} clusters during south-westerly winds (Fig. \ref{fig:SAR_panels} (b)). On the contrary, with easterly winds, (Fig. \ref{fig:SAR_panels} (c-d)), a streak of increased NRCS could be observed at the southern edge of the wake of several wind farms and wind farm clusters. 
Generalizing the observation, a streak of higher NRCS can be observed on the \emph{left side} of the wake region when moving downstream in the Northern Hemisphere.
 
Given that NRCS is related to $u^*$, it is also customary used to derive a wind speed at 10\,m above the sea surface, usually assuming a neutral boundary layer below 10\,m \citep[c.f][]{Djath.2019,Platis.2018, Hasager.2024}. In boundary layers, the friction velocity is canonically related to the shear stress, and hence governed by the covariance of the velocity fluctuations perpendicular to the wall, $u'w'$ and $v'w'$,
where $u'$, $v'$, $w'$ are the velocity vector components turbulent fluctuations. From the SAR point of view only, it is not clear whether the streaks of higher NRCS should be interpreted as an acceleration of the mean wind speed near the ground that locally triggers higher shear stress at the sea surface or as an increased level of Reynolds stresses and turbulence. Identifying the nature of such flow streaks is essential to frame how relevant they are to the recovery of the cluster wake, to the modification of the local climate, and to the potential impact on downstream wind farms. In fact, while an increased wind speed will only be beneficial for a downstream wind farm, a higher TKE will not.
On the one hand, a higher TKE at the inflow increases fatigue loads \citep{thomsen1999fatigue}, decreasing turbines' life expectancy. On the other hand, it promotes wake recovery \citep{Stevens2017} by enhancing the vertical turbulent entrainment, the fundamental mechanism of wake recovery, especially inside a wind farm operating in a boundary layer \citep{Calaf2010}.

An asymmetric behaviour in wind farm wakes has not only been observed through SAR. \citet{Platis.2018, Platis.2020} investigated some of the several airborne measurements of the wake of the N4 cluster in the German Bight collected by \cite{barfuss2019}. In one particular situation (September 10, 2016 08:30–09:30 UTC), they found a streak of increased turbulence kinetic energy (TKE) only on the western side of the wake region. 
This streak resembles the one in the NRCS of SAR because it also propagates more than 50\,km downstream of the wind farm cluster causing the wake, and it also appears on the left side of the cluster wake when moving downstream (the flight was conducted during southerly wind conditions). However, a very large horizontal velocity gradient (positive from left to right of the wake along the flow direction) was present in the undisturbed free stream  
In single turbine aerodynamics, such a situation is known to result in a larger TKE at the side of the wake neighbouring the fastest free stream. \citet{Chamorro2010, WuAndPorteAgel2012} show that a higher TKE is found at the top part of the wake of a turbine operating in a sheared boundary layer inflow. 
The observation of \cite{Platis.2018} by itself does not conclusively demonstrate the nature of the streaks observed in SAR for two reasons: the observations in SAR often depict a rather homogeneous wind speed across the wind farm wake, exhibiting an asymmetric behaviour; the TKE streak found by \citet{Platis.2018} is sampled at hub height and not near the ground.\\
Overall the characterisation of the turbulence in the wake of large offshore wind farm clusters via in-situ measurements is limited. However, significantly more understanding has been achieved through numerical studies.
Specifically, the work of \citet{Maas2022} and \citet{Lanzilao_Meyers_2025}, relying on Large Eddy Simulations (LES), provide evidence that a streak of higher TKE can form at only one side of the wake (always left along the flow direction in the northern hemisphere) also in the absence of a background flow gradient perpendicular to the main wind direction. None of these papers investigated further the physics of such a streak. However, the fact that a TKE streak appears in two different models based on different numerics is a hint that the effect is not a numerical artifact but rather a fluid dynamic phenomenon of the resolved physics.

Explaining the origin of the asymmetric TKE distribution in the wake of large wind farms is challenging. 
It is well known that inside a wind farm, the single turbine wakes are sources of TKE through production in the shear layer and decay of the tip vortices. 
The turbulence added by the single turbines is advected downstream in the wind farm wake. As proof, \citet{Djath.2018} found that under stable atmospheric conditions (probably overlapping with larger velocity vertical shear) the friction velocity may be increased with respect to the surrounding flow in the region 10\,km downstream of the wind farms in the German Bight due to larger vertical transport induced by the wind farm added turbulence.
However, the TKE added by the wind farm is expected to dissipate faster than the wake velocity deficit and to be reduced to a level lower than that in the free stream. This expectation is confirmed not only by the single observation of \citet{Platis.2018} but also by the long-term study on the loads of turbines exposed to a wind farm cluster wake of \citet{Anantharaman2025}. The authors, comparing loads at several turbines at the first row of a wind farm partially hit by a cluster wake, show that the turbines in the cluster wake have 7\% lower loads than their counterparts in the free stream, and only a 2\% increase in fatigue loads. This can be interpreted by assuming that the cluster wake reaching the sampling turbines has a lower absolute velocity than, and a similar TI to, the free stream outside the cluster wake. Thus, the TKE in the cluster wake must be lower than in the free stream. 
This phenomenon, also observed in LES \citep[c.f.][]{Maas2022, Lanzilao_Meyers_2025}, can be explained by the reduction of the vertical velocity shear in the cluster wake - the result of the momentum extraction of the turbines and the large mixing promoted by the added turbulence - which implies a loss of TKE production.

Overall, the process just described lacks a clear symmetry-breaking feature, i.e. an evident phenomenon that could justify the appearance of the asymmetric behaviour. Thus, the nature of the TKE streak must be searched elsewhere.
When looking at the physics of offshore wind farms, the factors that could induce asymmetries are the wind farm layout, the wind turbine rotor rotation direction, and the effect of the Coriolis force.
In atmospheric science literature, several papers have investigated how the Coriolis effect influences the interactions between an ABL and a hill. Among those, the work of \citet{Hunt2001} and \citet{Grisogno2004} show that when the Coriolis effect is not negligible, the velocity at the left and right sides of the hill wake has different values. This result was mainly found when the ABL is characterised by a subcritical Froude number, $Fr < 1$ ($Fr = \textrm{U} / N H$ with $N$ the Brunt-Vaisala frequency, $\textrm{U}$ the flow bulk velocity, and $H$ the inversion layer height). In such conditions, the flow is primarily displaced to the flanks of the hill as the vertical displacement above the topography is impeded by the buoyancy forces arising in the stably stratified free atmosphere. The lateral deflection of the flow imposed by the hill resulted in a speed-up on the northern flank and a slow-down at the southern flank (in the Northern Hemisphere and with westerly winds), which result from the Coriolis force.
While the analogy of a wind farm as a topographic hill has been proven reasonable for the study and modelling of the Global Blockage Effect \citep{Smith2010, Allaerts_Meyers_2019}, it is unclear if it may help understanding the formation of the streaks observed in LES or SAR. In fact, there is a large difference between the size of the hill considered by \citet{Grisogno2004} and the one of offshore wind farms. Furthermore, wind farms are porous obstacles, thereby inducing a lower flow displacement than a topographic hill.

\begin{figure}
    \centering
    \includegraphics[width=.9\linewidth]{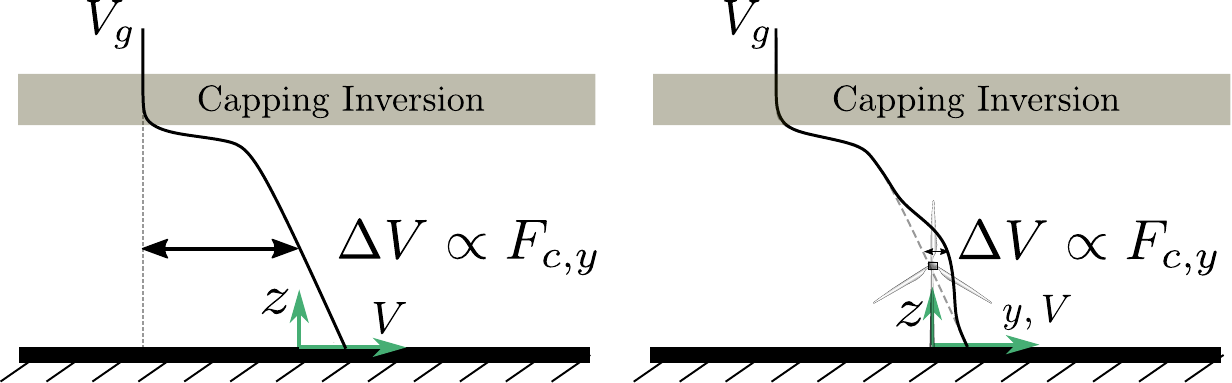}
    \caption{Schematics of how the Coriolis force in the crosswise direction $F_{\textrm{c},y}$ induces changes in the crosswise velocity $V$.  Panel (a) represents the veer present at the wind farm or turbine inflow that is the direct consequence of the Coriolis force arising from the boundary layer deceleration of an undisturbed streamwise velocity $U_\textrm{g}$.
    Panel (b) showcase the further change in $V$ localized in the turbine wake, again imposed by the Coriolis force arising from the streamwise velocity deficit inside the wake region.}
   
    \label{fig:coriosketch}
\end{figure}

The Coriolis force is also known to have an impact on the physics of single turbine wakes, as found by \cite{Abkar2016} and \cite{Vollmer2016}. For better discussion of their results, it helps to clarify that the Coriolis effect must be interpreted as the combination of two separate physical mechanisms \citep{vanderLaan.2017}. First, the presence of veer in the undisturbed wind (\autoref{fig:coriosketch} (a)). This results from the vertical gradient in the streamwise velocity inside the atmospheric boundary layer, i.e. shear. When the Coriolis force is present, the shear is associated with a wind direction that changes with height, known as veer or Ekman spiral. While in the surface layer (roughly the bottom 10\% of the ABL height) the large TKE promotes mixing of the veer, outside the surface layer, veer can no longer be neglected. Thus, in a relatively shallow atmospheric boundary layer, 200–600\,m high, veer characterizes the incoming flow for a wind farm equipped with modern turbine technology. The second Coriolis effect is a flow deflection, change of $V$, localized in the wake that results from the change in streamwise velocity imposed by the wake itself (\autoref{fig:coriosketch} (b)). Hereafther, we will refer to this force as the wake-deflecting (w.d.) Coriolis force. 

\cite{Abkar2016} show that veer is responsible for causing a deformation in the wake. They also found that the presence of veer increases the TKE level in the ABL, favouring faster wake recovery. More recently, \cite{Heck_Howland_2025} further investigated the sensitivity of turbine wake dynamics to the Rossby number, $Ro = U_\text{g}/(\Omega D)$ (with $U_g$ the geostrophic wind, $\Omega$ Earth’s rotation angular velocity, and $D$ the turbine rotor diameter), concluding that a lower $Ro$ is associated with a more intense anti-clockwise wake deflection. However, $Ro$ is a good predictor of the deflection magnitude and direction only when the inflow has no veer. This underlines that veer and the w.d. Coriolis force corresponding to the wake deficit compete against each other in skewing and deflecting the wake.
This complicated balance has been investigated also at the scale of wind farms. Using Reynolds Averaged Navier-Stokes (RANS) simulations, \cite{vanderLaan.2017} found that the higher vertical turbulent transport found in the wake of wind farms entrains the different wind direction of the layers aloft in the ABL, causing a clockwise turning of the farm wake at hub-height. \cite{Gadde2019} confirmed the observation with LES and studied the effect at different ABL stability. More recently, \cite{Lanzilao_Meyers_2025} show that when the ABL is very high and the veer weak, then the farm wake can only be deflected anti-clockwise by the w.d. Coriolis force.

Concerning the asymmetric behaviour of the TKE of wind farm wakes observed in LES, it is worth mentioning that in the project X-Wakes 
\citep{X-wakes}, we already demonstrated that the position of the TKE streak in LES is dependent on the Hemisphere for which the simulations are actually carried out. A simulation in the Northern Hemisphere shows the streak to form at the left side of the farm wake, while in the Southern Hemisphere at the right side. This result clearly indicates that the Coriolis force is at least one of the fundamental agents in inducing the asymmetric behaviour.
It is also noteworthy that offshore wind farm wakes have been related to an asymmetric ocean response with upwelling and downwelling regions related to the Coriolis force \citep{brostrom_influence_2008}.

However, there are still several unknowns concerning the TKE streak found in LES. Specifically regarding the physics of its formation, its impact on the surroundings, and more generally if it is a property only of wind farms or a broader fluid dynamic phenomenon of obstacles in an atmospheric boundary layer.
Thus, we formulate the following research questions to define the knowledge gap we intend to address in the current study:
\begin{enumerate}
    \item Is the Coriolis effect solely responsible for the symmetry breaking, resulting in a streak of higher TKE at only one side of the wake?
    \item What is the relation between the contributions of veer and the Coriolis force that deflects the farm wake and the formation of the TKE streak in the wind farm wake?
    \item Which physical mechanism is responsible for the TKE streak formation?
    \item Does the presence of the TKE streak impact the recovery of the wind farm wake?
\end{enumerate}
To provide an answer to all these questions, we make use of LES simulations that we set up considering a statistic of the atmospheric conditions correlated to the appearance of NRCS streaks in SAR. This allows us also to discuss whether the NRCS streaks observed through SAR are related to the TKE streaks observed in LES.

The paper is structured as follows: we introduce SAR and the LES model in \S\ref{sec:SAR} and \S\ref{sec:LES}, respectively. We investigate the appearance of the streak in SAR to find the best conditions for the LES set-up in \S\ref{sec:SAR-results}.  
We detail the LES set-up and a new model that allows for the simulation of boundary layers without Coriolis effects while maintaining the vertical profiles of ABLs that do include Coriolis (\S\ref{sec:LES-setup}), and detail statistics collection (\S\ref{sec:stats}).
In the results section, we explore the streaks in the LES at the chosen wind farm (\S\ref{sec:identification}), and answer the first and second of our research questions in \S\ref{sec:corio}. Furthermore, we describe     the physical phenomena leading to the formation of such streaks, answering our third research question (\S\ref{sec:streak_nature}). In \S\ref{sec:wr}, we focus on the impact of the streak and Coriolis on wake recovery.
Finally, we discuss similarities between TKE streaks and NRCS streaks in SAR (\S\ref{sec:discussion}).

\section{Methodology}
The main purpose of our study is to justify how TKE streaks appear in LES. However, we also intend to investigate whether the NRCS streaks in SAR are associated with such TKE streaks. Thus, we must simulate similar conditions in LES to those that promoted the streaks in SAR.
Hence, this section is divided into two parts. Initially, we outline the SAR fundamental principles (\S \ref{sec:SAR}). Then, in \S\ref{sec:SAR-results}, we present the atmospheric conditions (derived by a reanalysis model) found to correlate with the appearance of an asymmetric behaviour in several situations collected by SAR when passing over the German Bight. This part is necessary to identify the more interesting atmospheric conditions to be simulated in LES.
In the second part, we describe the LES model used (\S\ref{sec:LES}) and motivate the LES set-up chosen to numerically investigate if asymmetric streaks appear in LES, what their nature is, and their impact on the wind farm wake recovery. Moreover, we propose a novel method to perform LES simulations with the same vertical velocity profiles but without the Coriolis effects, essential to better clarify the physics behind the asymmetry found in LES (\S\ref{sec:LES-setup}). Finally, an overview of the collection of flow statistics in our LES is given (\S\ref{sec:stats}).

\subsection{Synthetic Aperture Radar}\label{sec:SAR}

In this study, imagery from SAR is used to provide evidence of situations where an asymmetric behaviour in wind farm wakes appears. SAR satellites transmit microwave radar signals that propagate through the atmosphere, mostly unaffected by meteorological conditions. Once the signal reaches the sea surface, it is primarily subject to Bragg scattering. Thus, the returned radar signals are dominated by the energy contained in small surface waves on a centimetre-scale generated by wind forcing. In fact, NRCS $\sigma_0$ is primarily correlated with the friction velocity $u^*$, defined as
\begin{equation}\label{eq:ufriction_formal}
    u^* = \sqrt{\nu (\partial u/\partial z)_{z=0}},
\end{equation}
with $\nu$ the vertical molecular momentum diffusion \citep{jones1978radar}.
Previous studies have shown that the near-surface wind speed can be derived from the NRCS using an empirical geophysical model function (GMF) in combination with additional information on wind direction and satellite incidence angle \citep{Portabella.2002,Hersbach.2007,verhoef_cmod5}. GMFs are commonly tuned for neutral ABL conditions, and additional considerations are required when applying these models in offshore wind farms wake regions, where significant modifications of vertical profiles of velocity, temperature, and turbulent fluxes can be expected. 
To avoid these complications, we do not convert the NRCS into 10\,m wind speed, as done in previous studies \citep{Christiansen.2005, Owda2022, Djath.2018, Hasager.2024}, but use the NRCS as an indicator for asymmetric wake features directly. 
 
The SAR images presented in Section \ref{sec:SAR-results} were acquired by the C-band Sentinel-1A/B satellites between 2017 and 2020 and are provided by the Copernicus Program, the European Commission's Earth Observation initiative (ESA, 2000–2020). The SAR images, captured in Interferometric Wide (IW) swath mode, offer a spatial resolution of 20\,m, which is sufficient to identify individual turbine wakes as well as small-scale features within offshore wind farm wakes. 

The SAR system has proven highly valuable for analysing wind farm wakes, as it provides instantaneous snapshots of the near-surface wind field across wide areas that encompass multiple offshore wind farms and clusters. Its potential for wake detection has been demonstrated in several studies \citep[e.g.][]{Hasager.2015b}, and it has more recently been applied to identify cluster wakes \citep{Schneemann.2020}.

\subsection{SAR data analysis: Conditions promoting farm wake asymmetry}\label{sec:SAR-results}
\begin{figure}
    \centering
    \includegraphics[scale=.8]{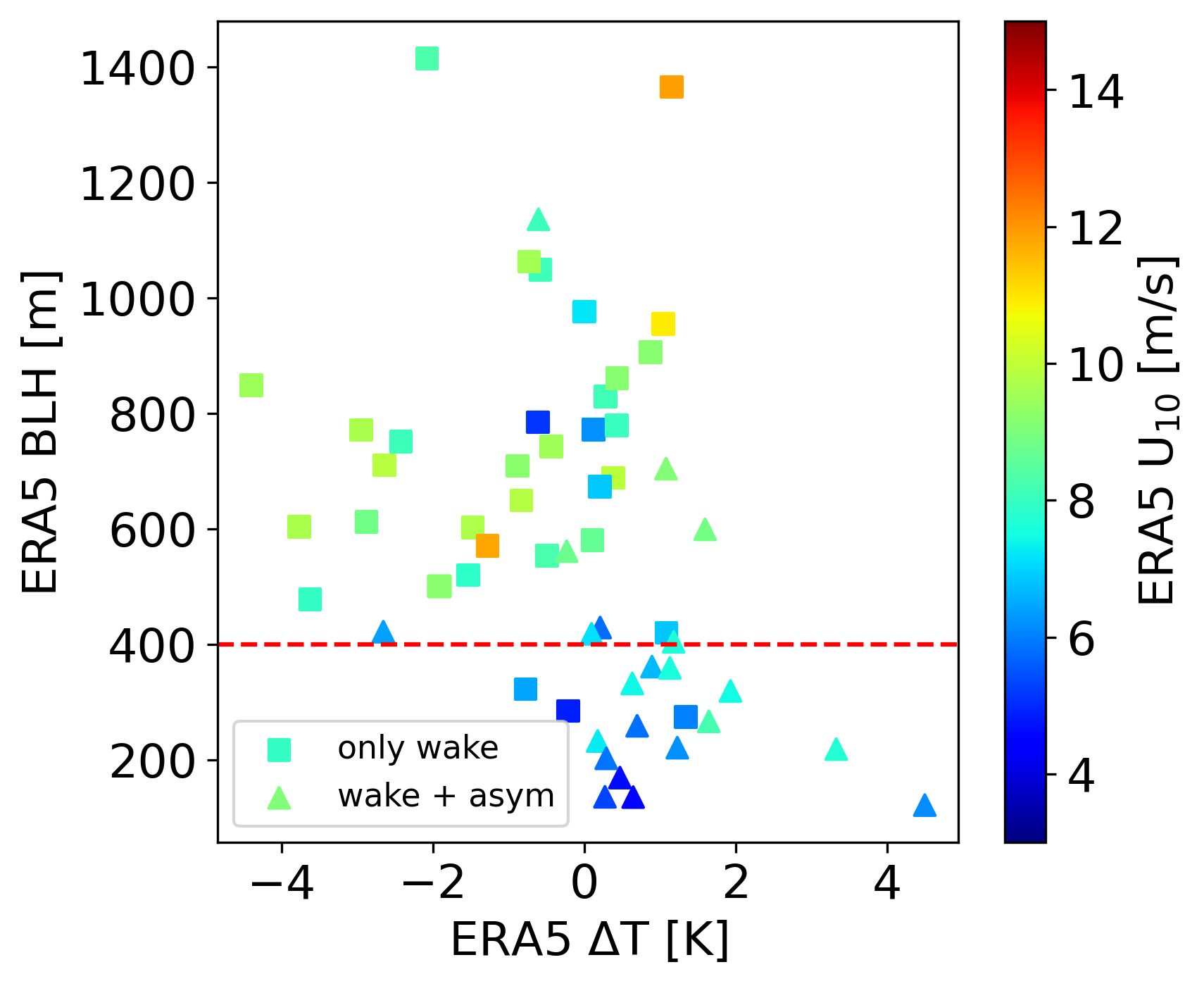}
    \caption{Occurrence of asymmetric cluster wake behaviour in relation to atmospheric conditions. The scatterplot shows boundary layer height and air/sea-surface temperature differences, both derived from ERA5, for colocated SAR cases in which wakes were observed behind offshore wind farms in the German Bight during 2018 and 2019. Marker colour indicates the velocity at 10\,m as calculated from ERA5. Cases exhibiting asymmetric behaviour, defined as a streak of enhanced NRCS confined to the left side of the wake relative to the wind direction, are indicated by triangular symbols.}
    \label{fig:SAR_parameter-map}
\end{figure}
We investigated a data set of SAR scenes collected over a two-year period to derive statistics on asymmetric wake features downstream of offshore wind farms in the German Bight.
The analysed situations were cross-correlated with the atmospheric stability, ABL height, and wind speed at the same time the SAR scenes were collected. 
We derived these atmospheric data from ERA5, the global atmospheric reanalysis produced by the European Centre for Medium-Range Weather Forecasts (ECMWF). ERA5 combines a numerical weather prediction model with a vast range of assimilated observations to provide a physically consistent estimate of the global atmosphere at hourly resolution and a horizontal grid spacing of about 31 km \citep{Hersbach.2020}.
The presented ERA5 data correspond to the grid point closest to the position of the offshore metmast FINO-1 (54$^\circ$01'N 06$^\circ$35'E). This corresponds to one of the most investigated measurement locations to characterise the German Bight atmospheric conditions, as it is located next to the first German offshore wind farm Alpha Ventus. 
In a first iteration, a subset of radar scenes showing offshore wind farms wake patterns was identified by visual inspection.  
The corresponding ERA-5 ABL heights, wind speeds at 10\,m, and air/sea-surface temperature differences (used as a proxy for atmospheric boundary layer stability) are shown in Fig. \ref{fig:SAR_parameter-map}.  
Among the SAR scenes displaying wakes, 47\% exhibit asymmetric behaviour, characterised by a narrow band of increased NRCS values, consistently located on the left side of the wake looking downstream. These cases are marked by triangle symbols in Fig. \ref{fig:SAR_parameter-map}. 
The cases showing an asymmetry correlate well with shallow atmospheric boundary layers, stable or near-neutral conditions (null or positive air/sea-surface temperature differences), and to a lesser extent with low velocity near the sea surface.
Such conditions are known to normally favour stronger vertical shear $\partial U / \partial z$ further away from the surface \citep{Stull1988}. 

However, further understanding of the mechanism that leads to the formation of the higher NRCS streak and whether the streak stems from an increase in the velocity or in the turbulence kinetic energy cannot be accomplished based on the SAR data alone.
Therefore, we must change the investigation framework from SAR to LES. 
The next section describes the numerical model and the setup choices taken.

\subsection{The LES numerical model}\label{sec:LES}
In this study, we rely on the Parallelized Large-eddy Simulation Model (PALM) v23.10, as defined by \citep{Maronga2020}. PALM is an open-source toolbox offering a very convenient solution for performing LES of the atmospheric boundary layer. It has been successfully applied to the field of wind energy, from wakes of single turbines and small wind farms \citep{Witha2014, Vollmer2016} to the scale of large offshore wind farms and wind farm clusters \citep{Maas_Raasch2022, Maas2022, Canadillas2023}. The interested reader can find examples where PALM was found to agree well with atmospheric data and wind tunnel measurements in \citet{Gryschka2014} and \citet{Gronemeier2021}, respectively. PALM was also found to perform well against other codes in simulating stable boundary layers by \cite{Beare2006}, and in simulating a wind turbine with an Actuator Line Model (ALM) in uniform inflow by \cite{Doubrawa2020}. Finally, \citet{centurelli2025} investigated the performance of PALM when applied to wind farm wakes modelling, the core topic of the current manuscript. The authors found PALM to agree well with a wind tunnel experiment on the mean wind speed and TI across the wake of a model wind farm with 24 actuator disks.
PALM solves the space-filtered non-hydrostatic Navier-Stokes equations in the Boussinesq approximation,

\begin{equation}\label{eq:NS}
    \begin{aligned}
          \frac{\partial \tilde{u}_i}{\partial t} = 
    - \frac{\partial \tilde{u}_i \tilde{u}_j}{\partial x_j}
    - \varepsilon_{ikm} f_k \tilde{u}_m
    + \varepsilon_{i3k} f_3 U_{g,k}
    - \frac{\partial}{\partial x_j} &\left( \frac{1}{\rho_0}\pi^*\delta_{ij} + \TAUIJ \right)\\
    &+ g \frac{\tilde{\theta} - \langle \tilde{\theta} \rangle}{\langle \tilde{\theta} \rangle} \delta_{i3}
    + d_i + S,
    \end{aligned}
\end{equation}
coupled with mass conservation equation, 
\begin{equation}\label{eq:conti}
\frac{\partial \tilde{u}_i}{\partial x_i} = 0,   
\end{equation}
and the filtered transport equation for temperature,
\begin{equation}\label{eq:energy}
\frac{\partial \tilde{\theta}}{\partial t} = -\frac{\partial (\tilde{u}_j \tilde{\theta})}{\partial x_j} - \frac{\partial}{\partial x_j} (\widetilde{u_j^{SGS} \theta_{\vphantom{j}}^{SGS}}),
\end{equation}
with $\tilde{\cdot}$ identifying filtered quantities, $f_i = [0,\, 2\Omega \cos{\Phi},\, 2\Omega\sin{\Phi}]$ the Coriolis parameter, depending on Earth’s rotational speed $\Omega$ and latitude $\Phi$, and $g$ gravitational acceleration.
The pressure term $\pi^* = p^* + \frac{2}{3} e \delta_{ij}$ contains the pressure perturbation $p^*$ and the trace of the sub-grid scale stress tensor, $e = \frac{1}{2} u^{SGS}_iu^{SGS}_i$, i.e. the kinetic energy of the unresolved scales. 
The perturbation pressure, $p^*$, is preferred to real pressure, $p$, as it represents only the pressure field that keeps the velocity divergence-free.
The forcing driving the flow is provided by the Geostrophic wind pressure gradient: $-\partial p / \partial x_i = 
\rho \varepsilon_{i3k} f_3 U_{\textrm{g},k}$.\\
The deviatoric part of the sub-grid scale stress tensor $\TAUIJ = \overline{u_i^{SGS} u_j^{SGS}} - \frac{2}{3} e \delta_{ij}$ is computed according to the Boussinesq hypothesis of proportionality to resolved scale gradients. The sub-grid scale eddy diffusivities for momentum and temperature are determined by the closure of \citet{Deardorff1980}, further modified by \cite{Moeng1988} and \cite{saiki2000}, and it is based on the solution of the prognostic equation for $e$.
Finally, $d_i$ is the turbine momentum sink and $S$ an additional source term, detailed in Tab. \ref{tab:cases}, that allows for simulations where Coriolis effects are manipulated (better clarified in \ref{sec:LES-setup}). 

Using finite difference schemes for the discretization of the model equations, PALM does not require the use of cyclic boundary conditions. 
Hence, unlike spectral codes, there is no requirement for fringe regions to remove wind farm-caused flow perturbations that reach the outlet, as in \citep{Stevens2014, Lanzilao2023}. 
Forward time marching is achieved with a third-order Runge-Kutta scheme \citep{Williamson1980}, and the fifth-order scheme of \citet{Skamarock2002} is used for the advection term.
The atmospheric simulations that PALM targets cannot be performed on Reynolds-dependent mesh resolutions due to the prohibitive computational cost. Therefore, it is customary that the first grid point above the no-slip bottom surface is located far outside the inner layer. Thus, a wall model is required to derive turbulence fluxes near the wall. As typically done for atmospheric boundary layer simulations \citep[c.f.][]{Moeng1984}, the Monin-Obukhov Similarity Theory (MOST) is used to model momentum, heat, and other scalar fluxes at the domain bottom surface.

For the remainder of the manuscript, we will drop the notation $\tilde{\cdot}$ used in Equations \ref{eq:NS}-\ref{eq:energy}. The filtered quantities resolved by LES will also be presented decomposed into mean and turbulent fluctuations according to the Reynolds' decomposition:
$u = U + u'$,
where $u'$ is only the resolved part of the fluctuating component of velocity.

\subsection{LES set-up for the investigation of the streak.}\label{sec:LES-setup}

LES of the interaction between the marine atmospheric boundary layer and large wind farms are computationally extremely expensive, as they require very large domains to limit the boundary effects on the solution and sufficient resolution to resolve the single turbines' rotors. Therefore, we decided to identify a single atmospheric scenario to be simulated for the investigation of the asymmetry in the wind farm wake. 
Considering the results shown in Fig. \ref{fig:SAR_parameter-map}, we decided to simulate a shallow Conventionally Neutral Boundary Layer (CNBL). 
Although asymmetric streaks tend to appear more consistently with stable stratifications, in LES, simulating a CNBL rather than a Stable Boundary Layer (SBL) allows us to resolve more of the turbulence spectrum at the same mesh resolution. In fact, as found by \citet{Beare2006}, SBLs are characterized by smaller turbulence scales and require higher computational effort to reach the same level of accuracy as a CNBL.
Furthermore, in the SAR analysis, we identified that the almost regular layout of the \emph{Amrumbank West} (ABW) wind farm operated by RWE in the N4 wind farm cluster of the German Bight produces streaks more consistently than other wind farms, especially when wind approached from West or East, as shown by SAR on the 21st of May 2018 at 5:48 UTC and the 13th of August 2020 at 5:49 UTC (Fig. \ref{fig:SAR_panels} (a, c)).
Therefore, we decided to simulate the ABW wind farm operating in purely westerly wind at hub-height with a velocity of 8\,m/s, i.e. in the turbines' partial load range.\\
The choice of simulating a single wind farm rather than a wind farm cluster stems from the need to limit the required domain size to favour a higher mesh resolution needed to resolve the turbulence properties investigated in sec. \ref{sec:results}.
\begin{figure}
    \centering
    \includegraphics[width=\linewidth]{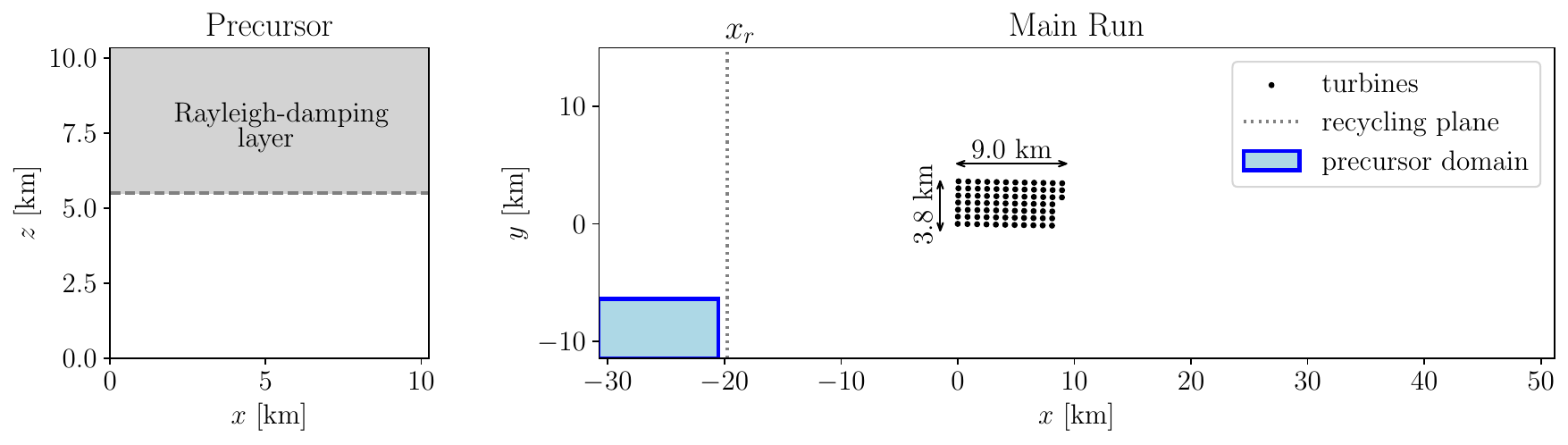}
    \caption{Precursor run vertical domain size (a). Main run horizontal domain size (b). Main run and precursor share the vertical direction. The region where the Rayleigh damping layer is applied is displayed as a grey shaded area in the precursor domain. In the main run, we showcase the precursor horizontal extent, the turbines, and the plane at which turbulent fluctuations are calculated to be recycled at the inlet.}
    \label{fig:domain}
\end{figure}

In PALM, the quasi-steady inflow conditions for the main simulations are achieved by means of a precursor simulation. 
The grid of the precursor simulation comprises 1024x504x256 points in the respective $x$, $y$, and $z$ coordinates. The spacing in the horizontal plane ($x$ and $y$) is constant and equal to 10\,m. In the vertical direction, a uniform spacing $\delta z = 10\,m$ is used only up to 500\,m; at larger heights, $\delta z_{k > 49} = \textbf{min}\left( 1.08 \delta z_{k -1},\, 50 \right)$\,m, with $k$ denoting the vertical index of the points of the computational grid. The vertical extent of the precursor domain, shown in \autoref{fig:domain} (a), is $\sim$10.3\,km.

To achieve a shallow boundary layer height of about 300\,m at the end of the precursor run, we prescribe an initial potential temperature profile with a capping inversion with a depth $\Delta H = 100$\,m that starts at a height of 200\,m and has a lapse rate $\Gamma = 2$ K/100\,m. In the free atmosphere aloft, the lapse rate is set to a value of $\Gamma_{\textrm{g}} = 0.35$ K/100\,m that is constant with height. This is the stratification of the ICAO standard atmosphere between the ground surface and the tropopause. \\
The geostrophic wind must be specified as it provides the volume force driving the flow and the initial condition of the precursor run, and the boundary condition at the top of the domain for all the simulations. Its streamwise and crosswise components are set to $U_\textrm{g} = 8.27$\,m/s and $V_\textrm{g} = -1.65$\,m/s, respectively. We do not use a wave model for describing the air/water interface; instead, we specify a constant roughness length, $z_0=2\cdot 10^{-4}$\,m, a value that is typical for offshore scenarios \citep[c.f.][]{Taylor2001}.

In the precursor, periodic boundary conditions apply at the streamwise $x$ and crosswise $y$ domain boundaries. As is the case for other codes, the application of streamwise periodic conditions causes the formation of unphysical infinite structures that span throughout the domain. To mitigate their impact on the solution, a shift in the $y$-direction, $\Delta y_{\text{shift}}$, is applied to the recycled absolute quantities from outlet to inlet (streamwise boundaries) as recommended by \citet{Munters2016}.

The development of the turbulent boundary layer in the precursor is known to cause gravity waves in the free atmosphere \citep{TAYLOR_SARKAR_2007} that could be reflected at the top boundary of the domain. To avoid the unphysical downward energy radiation associated with the reflection, PALM uses a Rayleigh damping layer (RDL) at the top of the domain, as suggested by \citet{Klemp1978}, that attenuates any perturbation before reaching the top boundary.
The same solution has to be adapted in the main run, where the wind farm imposes vertical motions that also trigger gravity waves \citep{Smith2010, Allaerts_Meyers_2017, Allaerts2018, Lanzilao_Meyers_2024}.
We adopted the same configurations for both precursor and main runs (as they share the vertical dimension $z$). The Rayleigh-damping layer is characterised by a viscous damping coefficient, $\nu$, that is sinusoidally increased from zero at the starting height of $z_{\text{RDL}}=5.5$\,km till 0.01 at the top of the domain. The layer has an overall depth of 4.8\,km. To guarantee that the RDL actually dampens upward propagating gravity waves (GW), it should be deeper than a gravity wave wavelength, $\lambda_\textrm{GW}$ \citep{Klemp1978, Khan2025}. 
To compute the vertical internal gravity wave length, we use the procedure proposed by \citet{Vosper2009} suitable for the heterogeneous stratification in the potential temperature profile.
\begin{equation*}
    \lambda_{\textrm{GW}} = 2\pi\frac{U_{\textrm{g}}}{ N_H}  \quad \text{with}\quad N_H =\sqrt{\frac{g}{\theta_0} \frac{\theta(H) -\theta_0}{H}}\quad\text{and}\quad H=H_{\textrm{ABL}} + \frac{U}{\sqrt{\frac{g}{\theta_0}\Gamma_\textrm{g}}}
\end{equation*}
The relation above predicts $\lambda_\textrm{GW} = 4.4$\,km. Since the RDL is deeper than the vertical wave length, it should be sufficient to attenuate GW reflection at the domain top boundary.

Turbulence in the precursor run is triggered by random fluctuations prescribed every 150\,s until the instantaneous TKE of the flow, calculated with respect to the horizontally averaged velocity vertical profile, is lower than a set threshold. The precursor simulation is run until a quasi-steady boundary layer is obtained, meaning that the TKE production is in balance with the computed mean velocity gradients. The volume-averaged TKE reaches a converged state after 45\,h of simulated time. \\
All prognostic quantities of the final integration timestep are saved, as they are used as initial conditions for the main run. The spatially averaged mean vertical profiles collected in the last hour of the precursor simulation are also collected to specify the mean inflow boundary condition in the main run. \\
The converged inflow boundary layer has a height of 320\,m, and a hub-height horizontal velocity of 8\,m/s with only the easting, $x$, component. 

The turbines are represented in PALM as actuator disks with rotation (ADM-R) \citep{Maronga2020}. This model has already been used by \citet{Doerenkaemper2015, Maas_Raasch2022} to investigate wind farm wakes and by \citet{Vollmer2016, Sengers2022} for investigating single wake deflection induced by rotor yawing. Both these latter works highlight that wake rotation and veer compete in modulating the single turbine wake shape. Hence, the turbine rotation should not be neglected when investigating asymmetric behaviours at a wind farm wake level. For this reason, in our investigation, we prioritise the ADM-R model over simple actuator disks to account for wake rotation.
However, we were not able to access all the detailed wind turbine information required by the ADM-R to simulate the wind turbines installed in ABW (Siemens SWT-3.6–120). Therefore, we opted to simulate the ABW wind farm using technical parameters of the exemplary NREL 5MW wind turbine \citep{Jonkman2009}. 
The NREL 5MW turbine is geometrically very similar to the Siemens SWT-3.6–120, with a rotor diameter $D = 126$\,m and a hub-height $z_\textrm{hub} = 93$\,m. However, it has a larger specific thrust.\\
We are aware that such a choice has a direct impact on the wind farm wake velocity deficit. However, the purpose of this manuscript is to investigate a fundamental physical phenomenon and not to replicate reality in detail quantitatively. Furthermore, as it will be shown later in the results section, the extent of the wake velocity deficit may affect the TKE streak intensity but not its formation nor its physical justification.

The domain of the main runs is significantly larger than that of the precursor in the horizontal directions, consisting of 8192x2640x256 grid points. However, the grid spacing is maintained at the same value as in the precursor run.\\
The initial conditions for the prognostic variables in the main run are determined from the 3D field in the last timestep of the precursor run through a cyclic fill approach. The inflow boundary condition for all the prognostic variables other than the potential temperature is a Dirichlet BC updated at each timestep $\xi(t_i, z, y, x=0) = \langle\overline{\xi_{\text{precursor}}(z,y,x)}\rangle_{x,y} + (\xi(t_{i-1}, z,y + \Delta y_{\text{shift}},x_r) - \langle\xi(t_{i-1},z,y,x_r)\rangle_{y})$, with $\langle\cdot\rangle$ representing spatial average and $x_r$ the coordinate of a plane parallel to the inflow 11\,km downstream where fluctuations are calculated. The potential temperature is instead recycled as an absolute value from the same plane at distance $x_r$ from the inflow, to mitigate the triggering of gravity waves due to mismatch in the potential temperature profile between the inlet boundary condition and the computational domain. This method of sustaining turbulence in the main run, initially proposed by \citet{Lund1998}, removes the need to compute the inflow in a separate simulation. 
Similarly to the precursor, the values collected at the recycling plane are shifted in $y$ by $\Delta y_{\text{shift}}$ before being applied at the inflow to avoid the formation of unphysical infinitely long turbulent structures \citep{Munters2016}.

The wind farm is located $\sim$30\,km downstream of the inlet, and the domain is 9 times larger along $x$ and 7 times along $y$ than the wind farm length and width, respectively. Such large domains were found in a previous study \citep[c.f.][]{centurelli.2021} to be required in order to minimise boundary impacts on the solution. 

To verify if a relation of causality exists between the Coriolis effect on the flow and the TKE streak, we conducted several LES simulations.
\begin{table}
  \begin{center}
\def~{\hphantom{0}}
  \begin{tabular}{lccccc}
  \textbf{Case} 
  & $\Phi$ [°] 
  & $\Omega$ [rad$\cdot$s$^{-1}$] 
  & \Vin [m$\cdot$s$^{-1}$] 
  & $S$ [m/s$^2$] \\
\hline
\textbf{Precursor} 
  & 54.5 
  &$7.29 \times 10^{-5}$ 
  & $V_{\textrm{g}}$ 
  & n.a. \\
\textbf{Precursor SH} 
  & -54.5 
  & $7.29 \times 10^{-5}$ 
  & -$V_{\textrm{g}}$ 
  & n.a. \\
\textbf{NH} 
  & 54.5 
  & $7.29 \times 10^{-5}$
  & $V_{\text{precursor}}$ 
  & n.a. \\
\textbf{SH} 
  & -54.5 
  & $7.29 \times 10^{-5}$
  & $V_{\text{precursor}}$ 
  & n.a. \\
\textbf{yE-nC} 
  & n.a. 
  & 0
  & $V_{\text{precursor}}$
  & $\varepsilon_{i3k} f_3 (U_{k,\textrm{in}} - U_{k,g})$ \\
\textbf{nE-yC} 
  & n.a. 
  & $7.29 \times 10^{-5}$ 
  & 0 
  & $-|\varepsilon_{i3k}| f_3 (U_{k,\textrm{in}} - U_{k,g})$ \\
\textbf{nE-nC} 
  & n.a. 
  & 0 
  & 0 
  & $\varepsilon_{132} f_3 (U_{2,\textrm{in}} - U_{2,g})$ \\
  \end{tabular}
  \caption{Overview of the latitude ($\Phi$), earth rotational speed ($\Omega$), inlet mean crosswise velocity profile, and momentum equation source ($S$) characterising the Coriolis effect in the different simulations performed.}
  \label{tab:cases}
  \end{center}
\end{table}
All the cases and the necessary precursors are reported in \autoref{tab:cases}. We considered a simulation in the Northern Hemisphere (NH), the natural location of the wind farm, and a simulation in the Southern Hemisphere (SH). For these two scenarios, two separate precursor runs are necessary, as shown by \autoref{tab:cases}.
Furthermore, we added three simulations of a fictitious atmospheric boundary layer, based on the NH precursor, where the Coriolis effects, i.e. the Ekman spiral (veer) and the wake-deflecting Coriolis force, are removed singularly or altogether. 
In particular, we simulated a case where the veer in the inflow is sustained throughout the domain but no further Coriolis force arises from flow wind speed change (yE-nC); a case with no veer and with the w.d. Coriolis force (nE-yC); and a case with neither veer nor the w.d. Coriolis force (nE-nC).

The method used to achieve such simulations is, to our knowledge, novel. Therefore, it is briefly outlined in the following.\\
When solving for the precursor, a boundary layer develops, decelerating the initial condition $u_i(t=0) = U_{g,i}$. As a result, the Coriolis term in Eq. \ref{eq:NS},
\[- \varepsilon_{ijk} f_j u_k
+ \varepsilon_{i3j} f_3 U_{g,j},\]
induces a height-dependent wind direction within the boundary layer that is referred to as the Ekman spiral or veer (\autoref{fig:coriosketch}). At the end of the precursor, the boundary layer reaches a quasi-steady state, and the spatially averaged mean velocity profiles, $ U_{\text{precursor}}$ and  $V_{\text{precursor}}$, collected to be used as part of the inflow BC of the main run, are sustained by the 
\begin{equation}
\label{eq:corio_pressure}
- \varepsilon_{i3j} f_3 ( U_{\text{precursor}, j}
- U_{g,j})
\end{equation} term.
To remove Coriolis effects from the solution, both veer at the inflow (\Vin $\equiv V(z,y,x=0) = V_{\text{precursor}}$) and the Coriolis parameter $f_3$ must be removed, i.e. set to 0. As a direct consequence, the forcing of Eq. \ref{eq:corio_pressure} cancels out and the mean velocity profile is no longer sustained throughout the main run domain. If the mean streamwise velocity profiles change across the simulations, their results are no longer comparable, as the interactions with the wind farm are seriously altered.

To ensure that the mean streamwise velocity profile developed in the precursor simulation is preserved in all the main run simulations where Coriolis effects are removed, we necessitate adding the source term $S$ in Eq. \ref{eq:NS}.\\
For the three different simulations yE-nC, nE-yC, and nE-nC, different sources are required. \autoref{tab:cases} summarizes the choices of $S$ and $\Omega$, affecting $f_3$, and the inflow mean veer $V(z,y,x=0)$ to achieve the suppression of one or both the manifestations of the Coriolis effect on the flow, namely veer and the w.d. Coriolis force induced by slow down with respect to the fully developed mean vertical velocity profiles of the ABL. These sources are defined to guarantee that the mean flow solution across these main runs does not deviate from the one found in the NH simulation.
\begin{figure}
    \centering
    \includegraphics[width=\linewidth]{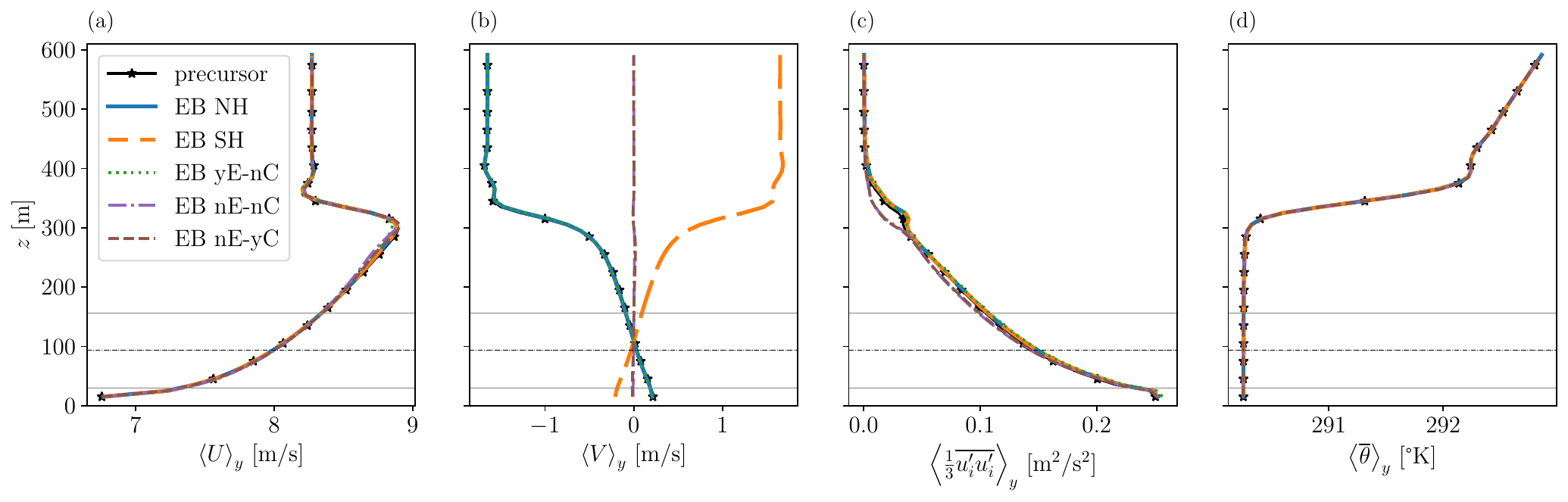}
    \caption{Mean vertical profiles of streamwise velocity (a), crosswise velocity (b), TKE (c), and potential temperature (d) averaged across the crosswise direction for the empty box main run simulation at $x=30$\,Km. These represents the inflow profiles to the wind farm in the 5 different simulation considered.
    The grey horizontal lines represent the turbine rotor vertical extent and the dash-dotted line the hub height.}
    \label{fig:inflow}
\end{figure}

To demonstrate how successfully the proposed method preserves the vertical profile of $U$, we performed empty box simulations (EB) for each of the main runs in \autoref{tab:cases}. These EB simulations are performed on a smaller domain with 3072x720x256 grid points but with the same grid spacing as the main runs and the precursor. They allow us to compute the development of the mean vertical profiles from the inlet boundary till the location of the wind farm first row.\\
\autoref{fig:inflow} displays the vertical profiles of the most relevant quantities at the outlet of the EB domain and the precursor for the NH simulation. The shear and the veer (for the simulations where it is not discarded) vertical profiles at the outlet of all the EB runs agree extremely well with the precursor simulation. We find the maximum deviation from the precursor, calculated as average in the ABL, to be less than 0.1\% for the nE-nC case. This shows that in all the considered boundary layers the streamwise development throughout the EB simulation domain is negligible. Assuming the same results hold in the case of the main run simulations, we can calculate quantities like the velocity deficit induced by the wind farm by normalising the velocity with the inflow profiles, avoiding a full empty box main run simulation for the same purpose. 

The vertical profiles of the TKE (Fig. \ref{fig:inflow} (c)) differ slightly across the simulations with no veer and the NH simulation. In fact, TKE reduces slightly across the domain in the nE-nC and nE-yC simulations, with a mean deviation in the ABL, calculated between inlet and outlet of the EB domain, of 4.3\% and 3.7\%, respectively. 
Although the TKE production in a boundary layer is dominated by the shear, $\partial U/ \partial z$, a small contribution also comes from veer, $\partial V / \partial z$. The lack of the latter causes the small differences in the TKE observed; however, given the overall qualitative comparisons across cases we performed in this manuscript, we deemed such differences negligible. Especially, as the mean velocity profile does not appear particularly modified by this small deficit in the TKE.
Another effect we neglected is the modification of the turbulence structures in the flow resulting from removing the Coriolis force (nC simulations), i.e. $f_i = 0$. Although this does not modify the mean velocity or TKE profiles in the free stream, it may have implications for single turbine wake recovery. However, further discussion in this direction is outside the scope of the current manuscript.

In the rest of the paper, we will refer to the simulations without veer/Ekman-spiral (nE) and without the w.d. Coriolis force (nC) as simulations without Coriolis effects. This is not entirely correct, as the total absence of Coriolis, achieved by developing a precursor also without Coriolis, would have resulted in a different shear profile than the one calculated in the NH simulations \citep{vanderLaan2021}.

\subsection{Statistics and Data Analysis}\label{sec:stats}

In \S\ref{sec:results}, we will present first and second-order flow statistics that require a sufficiently long averaging time before reaching convergence. For the main runs, a total simulated time of 6\,h is chosen, and statistics are only derived from data collected over the last 3\,h. This represents a full flow-through throughout the domain.
 
The collection of the second-order statistics and the TKE budgets is achieved by applying Reynolds decomposition to all the terms. Double and triple products of velocity can be derived as
\begin{subequations}
    \begin{equation}\label{eq:double_product}
        \overline{u'_iu'_j} = \overline{(u_i - U_i)(u_j - U_j)}=\overline{u_iu_j} - U_iU_j   
    \end{equation}
    \begin{equation}
        \begin{aligned}
        \overline{u'_iu'_ju'_k} = \overline{(u_i - U_i)(u_j - U_j)(u_k - U_k)} &= 
\overline{u_i u_j u_k} \\
&- \overline{u_i u_j} U_k
- \overline{u_i u_k} U_j
- \overline{u_j u_k} U_i
- U_i U_j U_k.
\end{aligned}
    \end{equation}
\end{subequations}
Any other term appearing in the TKE budget equation can be derived following Eq. \ref{eq:double_product}.

All the flow statistics are later further treated with a moving average with a window size of 100\,m in the horizontal directions only ($x$ and $y$) to attenuate the spurious turbulent fluctuations that the time averaging has not removed.

\section{Results}\label{sec:results}
This section is dedicated to presenting results that address our research questions.
After having set up our reference LES (NH) to replicate the conditions for which asymmetric behaviour in the wake of a wind farm appears according to SAR (\S\ref{sec:SAR-results}), we initially focus on showing in which field the numerical model identifies an asymmetry in \autoref{sec:identification}. 
Further, in \autoref{sec:corio}, we demonstrate which of the Coriolis effects is responsible for the asymmetry formation (research question 1). Moreover, in \autoref{sec:mechanism}, we clarify the physical mechanism that leads to the formation of the TKE streak in LES (research question 2). Finally, in \autoref{sec:wr}, we make considerations on the wake recovery connected to cases where the asymmetry is found (research question 3).
\subsection{Identification of the asymmetry through LES}\label{sec:identification}
\begin{figure}
    \centering
    \includegraphics[width=\linewidth]{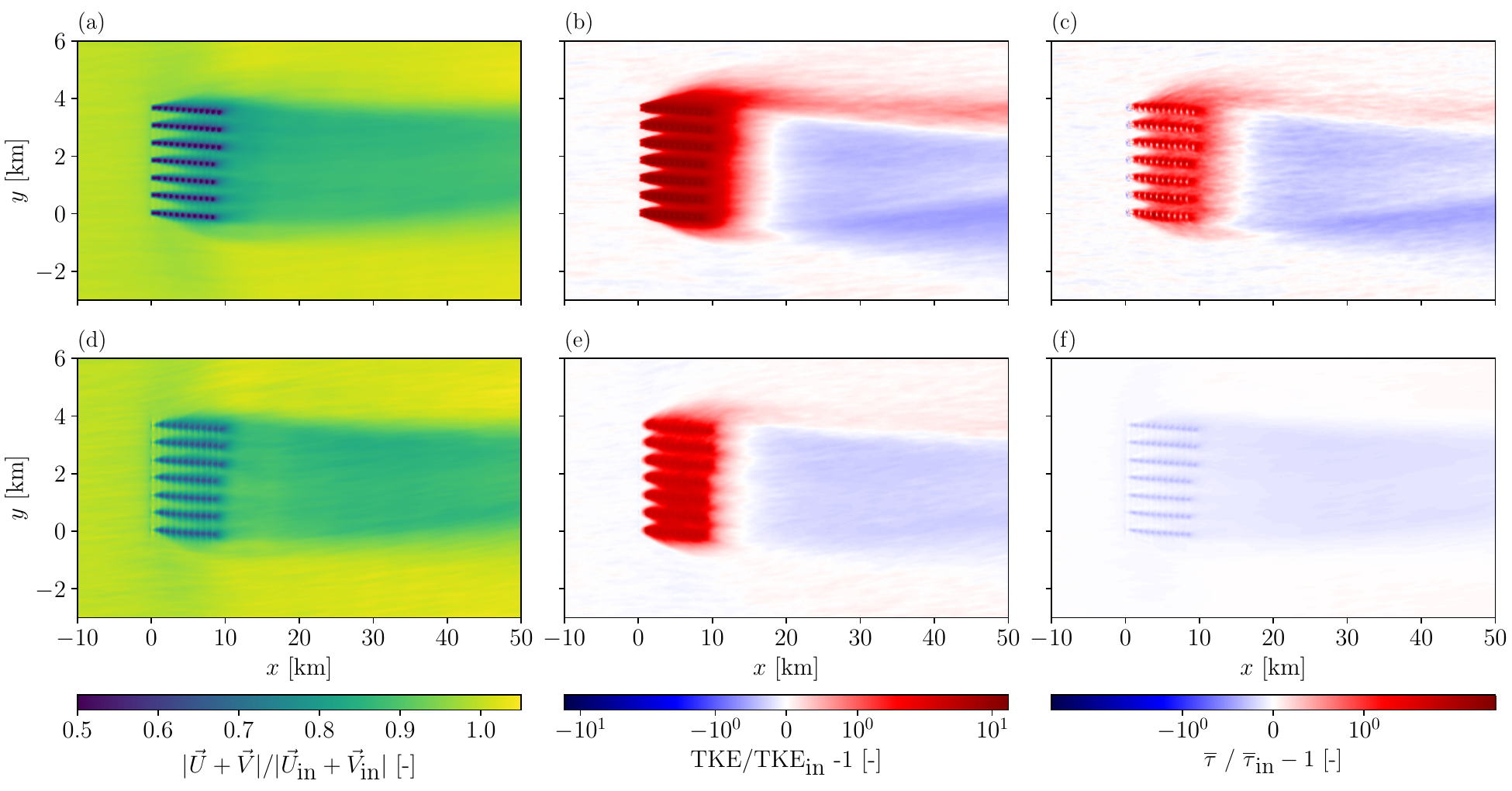}
    \caption{Normalised horizontal velocity (a,d), TKE perturbation (b,e), and horizontal shear stress ($\tau$) perturbation (c,f) for the NH simulation sampled at horizontal planes at hub-height (a-c) and 10\,m above the domain bottom surface (d-f). The results at 10 metres are interpolated from the first and second grid points from the ground. Note that the normalised TKE and $\tau$ fields are not expressed as percentages. The large extent of the colourbar is necessary to avoid clipping in the very near turbine wake. We used logarithmic scaling to appreciate the more modest increase of the two quantities in the streak region. Such increase is still in the order of 50\% (0.5 according to the colourbar) of the value in the free stream at hub-height.}
    \label{fig:streak_identification}
\end{figure}
To begin, we assess whether a streak similar to the one observed in the SAR imagery (Fig. \ref{fig:SAR_panels}) is also present in the LES results.\\
\autoref{fig:streak_identification} shows the normalised horizontal velocity, turbulence kinetic energy, and horizontal shear stress for the NH simulation. Panels (a) and (b) demonstrate that at hub-height, at the left edge of the wake along-the-flow-direction (f.d.), LES computes a streak of increased TKE but not of mean horizontal velocity. In particular, the TKE in the streak at hub-height is almost 50\% and 100\% larger than in the free stream and in the wake region, respectively. 
This distinct TKE behaviour has already been observed in other numerical studies, specifically \citet{Maas2022} and \citet{Lanzilao_Meyers_2025}.
In the velocity field, the wake does not appear particularly asymmetric; however, it is possible to observe the curvature in the wake propagation that could be justified by an initial clockwise deflection induced by veer mixing and later by the anti-clockwise deflection imposed by the Coriolis pressure gradient \citep{vanderLaan.2017,Lanzilao_Meyers_2025}. 

The analysis above focuses on hub-height quantities. However, later we will discuss whether a connection between the TKE streak in LES and the NRCS streaks observed by SAR at the sea surface exists. Therefore, we show also the values of normalised velocity and TKE at 10\,m above the domain bottom surface in Fig. \ref{fig:streak_identification} (d, e). Similarly to hub height, no asymmetric behaviour or streak is found in the velocity field, while a streak is still visible in the TKE field, although more attenuated.

According to the LES results presented, the streak measured by SAR could match with a higher TKE level found at only the left side of the farm wake (similar to \autoref{fig:SAR_panels} (a)).
However, formally, the NRCS measured by SAR is correlated with the friction velocity, $u^*$ \citep[c.f.][]{Djath.2018, Durden1985}, and not directly to the TKE.
In wall-modelled LES, as is the case for our simulations, resolution near the wall is not sufficient to calculate $u*$ according to \autoref{eq:ufriction_formal}. Therefore, we resort to the following formulation to estimate it.
\begin{equation}\label{eq:us}
    u^* = \sqrt{\frac{\overline{\tau_w}}{\rho}} = -\left.\left[\left(\overline{u'w'} + \tau^{\text{sgs}}_{13}\right)^{2}+ \left(\overline{v'w'}  + \tau^{\text{sgs}}_{23}\right)^{2}\right]^{\frac{1}{4}}\right|_{z=z_{\textrm{SAR}}} \equiv \tau(z=z_{\textrm{SAR}}),
\end{equation}
with $z_{\textrm{SAR}}$=10\,m, $\tau_w$ the shear stress on the bottom domain surface, and $\tau$ the generic horizontal shear stress in the fluid.
The equivalence above is well justified by the fact that in the surface layer of the ABL the total shear stress is constant \citep{Stull1988}.
Since we substitute the calculation of $u^*$ with the generic mean shear stress $\tau$, we can then display its perturbation field both at hub-height and at 10\,m ($\propto$\,$\tau_w$) in \autoref{fig:streak_identification} (c, f). 
At hub-height, $\tau$ behaves similarly to the TKE. However, near the ground, the streak in the field of the horizontal shear stress on the northern side is rather faint and becomes indistinguishable after $\sim$10\,km. This contrasting result is examined further in \autoref{sec:discussion}. 

With this section, we highlight that LES suggests that a streak with a significantly higher ($\sim$50\%) TKE than the free stream can be generated at the left side f.d. of the wake. This bears important consequences on the farm wake itself, but more importantly on hypothetical downstream wind farms. In the next sections, we deepen the understanding of the formation of such features. 
\subsection{Role of veer and wake deflecting Coriolis force}\label{sec:corio}
\begin{figure}[]
    \centering
    \includegraphics[width=\linewidth]{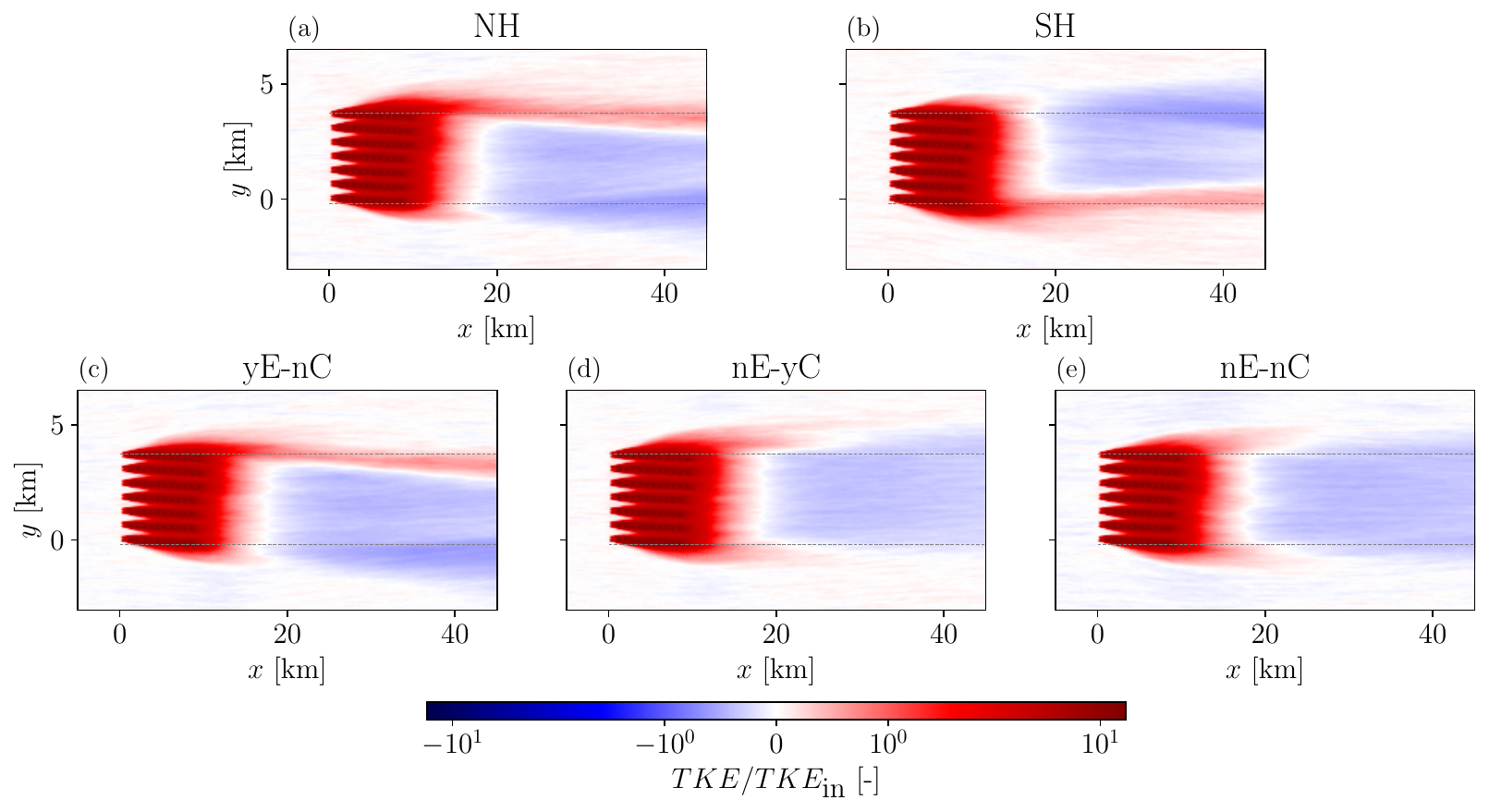}
    \caption{Mean hub-height TKE perturbation for all the simulations performed: NH (a), SH (b), yE-nC (c), nE-yC (d), and nE-nC (e). The streak switches side when the hemisphere is changed, coherently with the observations in SAR (always in the northern hemisphere but with westerly and easterly winds).
    The streak is very evident only in the cases where the Ekman spiral (veer) is present: NH (a), SH (b), and yE-nC (c). Only the w.d. Coriolis force resulting from the deceleration induced by the wind farm wake, case nE-yC (d), induces only a minor asymmetry that is not sustained throughout the computational domain. The latter case is similar to the case where Coriolis is completely absent: nE-nC (e).
    Horizontal grey lines indicate the upwind/downwind projections of the wind farm lateral edges, helping to visualise the wake deflection out of the imaginary corridor of a perfect mean flow advection.}
\label{fig:TKE_coriolis}
\end{figure}

In this subsection, we investigate the relation between the Coriolis force and the TKE streak found in Fig. \ref{fig:streak_identification} (b,e).\\
\autoref{fig:TKE_coriolis} displays the TKE perturbation field at hub height for the five main run simulations performed.
We start by considering only the results of the NH (already shown before) and the SH simulation (Fig. \ref{fig:TKE_coriolis} (a, b)). The switch in hemisphere results in the streak changing side from the left to the right side of the wake, confirming the results found in \citet{X-wakes}. It is important to note that the simulation set-up of NH and SH is identical, except for an antisymmetric value of the latitude that results in an antisymmetric distribution of veer (from the precursor) and w.d. Coriolis force.
The wind farm layout is not mirrored between NH and SH simulations. Therefore, it becomes clear that the slightly asymmetric ABW layout – specifically, the presence of three additional turbines only in the northern half of the farm, in the last row – has no impact on the streak formation. \\
The final proof that the streak formation is rooted in the Coriolis effect is offered by the nE-nC simulation.  
When fully removing both veer and the Coriolis force deflecting the wake, the behaviour of the TKE in the wind farm wake is symmetric and no clear streak survives past $\sim$12\,km downstream of the wind farm. 

While it is clear that the Coriolis effects are involved in the TKE streak, it is not clear yet whether veer, the w.d. Coriolis force, or both are necessary for its formation.
Fig. \ref{fig:TKE_coriolis} (c, d) demonstrate that the TKE streak is mostly uniquely induced by veer. Only the w.d. Coriolis force is not sufficient to introduce a long-lasting asymmetry in the TKE of the wind farm wake region. In fact, while a higher TKE at the left wake edge survives longer than at the right wake edge, this asymmetry vanishes before the domain outlet.\\
The yE-nC and nE-yC cases also isolate the counter-acting effects of the two mechanisms through which the Coriolis effects deflect the wind farm wake. The lack of the w.d. Coriolis force (yE-nC) results in a downward or clockwise deflection induced by veer mixing, like in \citet{vanderLaan.2017}. The lack of veer (nE-yC) leads to a faint counter-clockwise rotation. As found by \citet{Lanzilao_Meyers_2025}, in a shallow boundary layer the veer mixing has greater importance than the w.d. Coriolis force in deflecting the wake, as the change of the wind direction with height ($z$) is stronger in such a case.
\subsection{Reasoning of the physics behind the streak formation}\label{sec:mechanism}
The previous two subsections show that the Coriolis force, by inducing veer in the wind approaching the wind farm, is the main driver determining the appearance of the TKE streak. However, the physical mechanism behind the streak formation remains uncovered.
\begin{figure}
    \centering
    \includegraphics[width=0.9\linewidth]{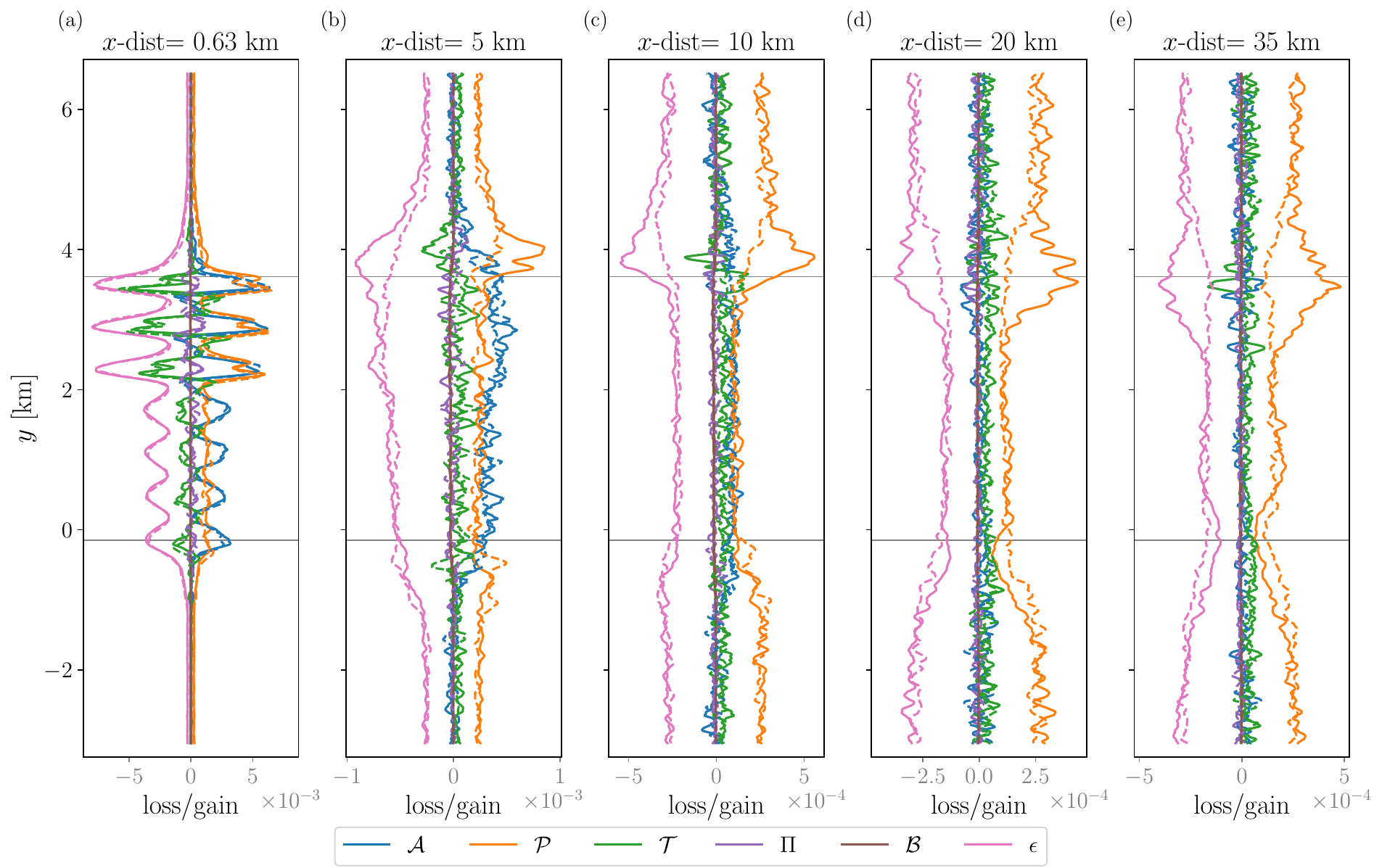}
    \caption{TKE budgets, as defined in \autoref{eq:tke_bud}, at hub height plotted along the crosswise direction $y$ for different distances downstream of the last turbine of the wind farm ($x$-dist). Solid lines (\rule[0.5ex]{.5cm}{0.5mm}) represent the NH simulation results and dashed lines (\hdashrule[0.5ex]{0.7cm}{0.5mm}{1mm}) the nE-nC. Horizontal grey lines mark the wind farm extent along $y$.}
    \label{fig:tke_budgets}
\end{figure}

We start by investigating the TKE budgets for the NH and the nE-nC simulations, to understand how the larger TKE is formed in the streak region.
The budget equation for the turbulence kinetic energy of the resolved scale, $k$, reads:
\newcommand{\vspacer}{\vphantom{\frac{1}{\varrho} \frac{\partial  (\overline{u'_i \pi^{*'}})}{\partial x_i}}}
\begin{equation}\label{eq:tke_bud}
    \begin{aligned}
     0=-\overbrace{\vspacer \UMEAN_j\frac{\partial \overline{k}}{\partial x_j}}^{\displaystyle \mathcal{A}} -
     \overbrace{\vspacer \frac{\partial}{\partial x_j} \left(
     \frac{1}{2} \overline{u'_i u'_i u'_j} + \overline{u'_i \TAUIJ}\right)}^{\displaystyle \mathcal{T}} -  
    \overbrace{\vspacer \frac{1}{\varrho} \frac{\partial  (\overline{u'_i \pi^{*'}})}{\partial x_i}}^{\displaystyle \Pi}
      - \overbrace{ \vspacer \overline{u'_i u'_j}\overline{S\vphantom{u'_i}}_{ij}}^{\displaystyle \mathcal{P}} &+ \overbrace{g\,\overline{u'_3 \left(\frac{\vartheta - \thetaref}{\thetaref} \right)'}}^{\displaystyle \mathcal{B}}  +\\
     &  + \underbrace{\vspacer \overline{\tau^r_{ij} S'_{ij}} + \mathcal{R}}_{\displaystyle \epsilon}.  
    \end{aligned}  
\end{equation}
We assume the time average we use is sufficient to guarantee $\frac{\partial k}{\partial t} =0$.
The different terms in Eq. \ref{eq:tke_bud} stand for mean flow advection ($\mathcal{A}$), resolved and modelled turbulent transport ($\mathcal{T}$), turbulent pressure transport ($\displaystyle \Pi$), $k$ production ($\mathcal{P}$), buoyancy transport ($\mathcal{B}$), and dissipation ($\displaystyle \epsilon$). Note that the dissipation includes the numerical residual of the budgets ($\mathcal{R}$). This choice stems from the fact that the Wicker-Skamarock scheme \citep{Skamarock2002} used in the advection term discretisation is very dissipative. Hence, a large part of the residual is constituted by numerical dissipation not explicitly accounted for by the TKE budget equation, as shown by \citet{Heinze2015}.
Furthermore, we calculate the buoyancy term although it is expected to not be very relevant \citep{Baungaard2022}, because the potential temperature change inside the boundary layer, induced by turbulent entrainment from the capping inversion, is negligible. The Coriolis term does not directly affect the TKE but only distributes energy across its components. 

Figure \ref{fig:tke_budgets} shows the hub-height local TKE budgets along lines perpendicular to the inflow wind direction along $y$ for several downstream distances from the last turbine of the farm. We discuss the results in the following subsections.

\subsubsection{TKE budgets in the farm near-wake}
Borrowing the wording from single turbine wakes literature, we identify the near-wake region as the part of the farm wake where the TKE is strongly affected by single turbine physics ($\sim$2\,km downstream of the last turbines of the farm). According to Fig. \ref{fig:TKE_coriolis}, this region is not characterised by an evident asymmetry yet. Thus, it may seem that its analysis does not align with the research questions of this study.
On the contrary, we think that two main reasons exist to discuss the TKE budgets also in this region. On the one hand, since the single turbine contributions are still evident, we can compare our TKE budget results to those of established literature for single turbines. On the other hand, we can better verify that the farm wake behaviour later described does not originate from the superposition of single turbine physics. 

We start analysing panel (a) of \autoref{fig:tke_budgets}. The TKE budget horizontal profiles are plotted at a distance of 0.63\,km from the last turbine of the farm. The contributions of single turbine wakes to the TKE advection, dissipation, and turbulent transport are evident at this distance.
It is also possible to observe major differences in these budgets between the left and right f.d. halves of the farm wake (top and bottom halves in each subplot, respectively).
This is the result of the ABW wind farm layout, characterised by a last row with only three turbines located in the northern half of the farm.
This means that on the left half of the wake the distance from the last turbines of the farm is 0.63\,km, or $\approx$5\,D. The signature of the other four wakes in the right half of the wake is from the turbines in the penultimate row of the wind farm, located 1.45\,km or 11.7\,D upstream.

The sampled budgets at these two downstream locations, 5\,D and 12\,D, reveal distinct dynamics. At both positions, advection and production support the elevated TKE levels characteristic of turbine wakes \citep{pope2001, rind_castro2012}. However, while the TKE production is comparable to advection at 5\,D, it becomes significantly weaker at 12\,D. This is expected, as the strong turbulence of the flow experienced by turbines deep inside the farm quickly breaks down the shear layers that sustain production. Similarly, the turbulent transport, negative within the wake due to outward diffusion, also decays rapidly between 5\,D and 12\,D.

Overall, the TKE budgets of turbines deep in the farm sampled at these two locations cannot be rescaled to match each other, thus they seem to lack the self-similar property found by \citet{rind_castro2012}. Investigating further this behaviour is outside the scope of the current study. However, it can be expected that the mean flow shear of the ABL and the strong streamwise decay of the turbulence added by the wind farm, which are both missing in the homogeneous and laminar inflow used by \citet{rind_castro2012}, break the streamwise similarity. 
The very high TKE found in the near wind farm wake and its subsequent streamwise decay may also be responsible for the qualitative differences we found between our results for the single turbine wake at 5\,D and the one reported by \citet{Bastankhah.2024}. These authors conducted a study of TKE at model turbines in a boundary layer with a TI of only 4.8\%  in the inflow. Such a TI value, while very representative of the conditions at the first row of an offshore wind farm, is very low for waked turbines.
Finally, it is worth emphasizing that a direct comparison of our results to the experimental ones mentioned here is further hindered by differences in turbine thrust coefficients and turbine modelling through the ADM-R.

One of the main conclusions from this subsection is that the budgets across the NH and nE-nC simulations at 0.63\,km downstream of the farm do not differ significantly. This underlines that the TKE streak does not appear to originate from single turbine physics.

\subsubsection{TKE budgets in the farm far-wake}
When analysing the TKE budgets further downstream, the asymmetric behaviour for the NH case clearly emerges.
In fact, at 5\,km (\autoref{fig:tke_budgets} (b)), the TKE production at hub height in the farm wake is roughly the same as in the free stream; however, a spike of production is noticeable at the left f.d. edge of the wake for the NH case. The spike evolves further downstream, clearly separating the TKE production in the NH case from that in the nE-nC case. An opposite behaviour also appears on the southern side of the farm wake around $x-$dist = 20\,km. Here, the TKE production for the NH case exhibits a minimum.

While the dissipation has a behaviour that mirrors that of the TKE production, all the other budgets shown in Fig. \ref{fig:tke_budgets} are quite equal between the NH and nE-nC simulations. Only the TKE turbulent transport and advection show some spurious peaks in the high TKE streak, suggesting they play a minor role in redistributing the extra TKE produced in the NH case.
Similar conclusions could be drawn also when analysing the TKE budgets of the SH and yE-nC (not shown here).\\ 
From this result, we determine that the TKE streak is not the result of an accumulation process, whereby turbulent transport and/or advection collect TKE from elsewhere and transport it into the streak region. The larger TKE in the streak is uniquely due to a localized source mechanism connected to the TKE production.

Although the results of Fig. \ref{fig:TKE_coriolis} and \ref{fig:tke_budgets} show that veer produces the streak by inducing a larger TKE production, the mechanism through which this happens remains unclear. Thus, next we investigate deeper what causes the larger TKE production. 

\subsubsection{Origin of the larger TKE production causing the streak}\label{sec:origin}
\begin{figure}
    \centering
    \includegraphics[width=0.9\linewidth]{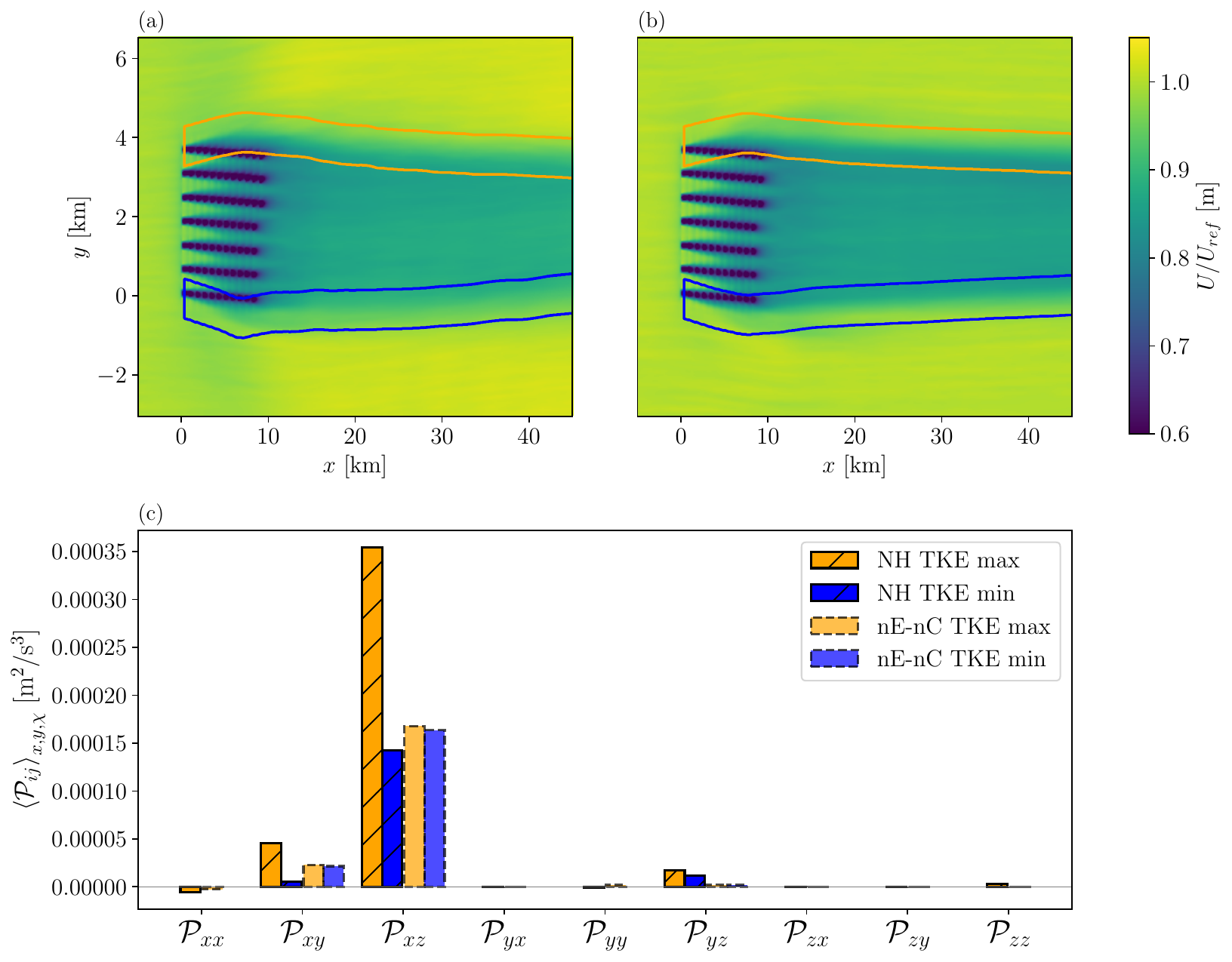}
    \caption{Isolation of the left and right f.d. wake side where TKE max and TKE min regions are defined for the NH case (a). The same algorithm based on velocity deficit is applied to the nE-nC simulation (b). Components of the TKE production according to \autoref{eq:production_components} averaged throughout the TKE max control volume (orange outline in panels (a) and (b)) and from 20 to 300\,m in the vertical direction $z$ for the NH and nE-nC simulation. For the most relevant terms we also display the same values at the TKE min region.}
    \label{fig:tke_production_bars}
\end{figure}
To further investigate how veer induces the larger TKE production at the left side of the wake in the NH simulation, we first split the production budget $\mathcal{P}$ into the several components $\mathcal{P}_{ij}$ corresponding to the mean flow deformation matrix $\partial U_i/\partial x_j$. 
Subsequently, we calculate these terms in a control volume corresponding to the flow region where the larger TKE production is found. 
The control volume is defined as a band of width $\mathcal{\Delta}y=1$\,km around the curvilinear coordinate $\chi_{\textrm{TKE max}}$ defined as the locus of points at the northern wake side where the hub-height velocity reaches 75\% of the free stream. This region is extruded vertically, between $z_{\textrm{min}} = 20$\,m and $z_{\textrm{max}} = 300$\,m, to close the control volume.
We chose to identify the region of the TKE streak from the velocity field as it aligns with the horizontal shear layer between the farm wake and free stream. Furthermore, a criterion based on the velocity is more versatile than one based on TKE, because the latter would not be applicable for the nE-nC case where no TKE streak appears. We needed to isolate the wake edges separately in the NH and nE-nC simulations as the wake propagates in slightly different directions in the two simulations. 
We show the results of the isolation of the left and right sides of the wake in Fig. \ref{fig:tke_production_bars} (a,b). For generality, since the results of the Northern Hemisphere (NH) simulations mirror those of the Southern Hemisphere (SH) but with opposite symmetry in terms of the wake side on which they occur, we will hereafter refer to the left side of the wake as the TKE max region and the right side as the TKE min region. This convention applies also to the nE-nC simulation, although no difference is observed in the TKE across the two sides of the wake.

Fig. \ref{fig:tke_production_bars} (c) displays the TKE production components averaged in the TKE max control volumes,
\begin{equation}\label{eq:production_components}
    \left \langle \mathcal{P}_{ij} \right \rangle_{x,y, \chi} = \frac{1}{\Delta y }\frac{1}{\Delta \chi}\frac{1}{(z_{\textrm{max}} - z_{\textrm{min}})}\int_{z_{\textrm{min}}}^{z_{\textrm{max}}}\int_{-0.5\mathcal{\Delta} y}^{0.5\mathcal{\Delta} y}\int_{\chi_{\textrm{min}}}^{\chi_{\textrm{max}}} \overline{u'_iu'_j} \frac{\partial U_i}{\partial x_j}\, dx\,dy\,d\chi_{\textrm{TKE max}},
\end{equation}
for the NH and nE-nC simulations.
In both cases, the TKE production is dominated by the term connected to the vertical shear $\mathcal{P}_{xz}$. The production from the streamwise velocity gradient in $y$, $\mathcal{P}_{xy}$, is also a minor contributor.
Only in the NH simulations, the presence of veer also induces a small but non-negligible contribution $P_{yz}$.\\
The results across the two wake edges (orange and blue bars in Fig. \ref{fig:tke_production_bars} (c)) are largely different for the NH simulation, while they are almost identical in the nE-nC simulation. This result expands on what was already presented by \autoref{fig:tke_budgets}, as it aggregates the production budgets across the whole ABL height.\\
From this analysis, it becomes evident that despite finding an asymmetric behaviour in both $P_{xz}$ and $P_{xy}$ for the NH simulation, quantitatively, a larger $P_{xz}$ is what produces the TKE streak in the NH case with respect to the nE-nC simulation. This statement is corroborated by Appendix \ref{sec:appendix1}. In particular, we show in \autoref{fig:tke_prod_components} that in the TKE max region for the NH simulation, while all the Reynolds stresses are homogeneously increased, specifically the $\partial U / \partial x$ is significantly larger than in the nE-nC simulation.

\begin{figure}
    \centering
    \includegraphics[width=\linewidth]{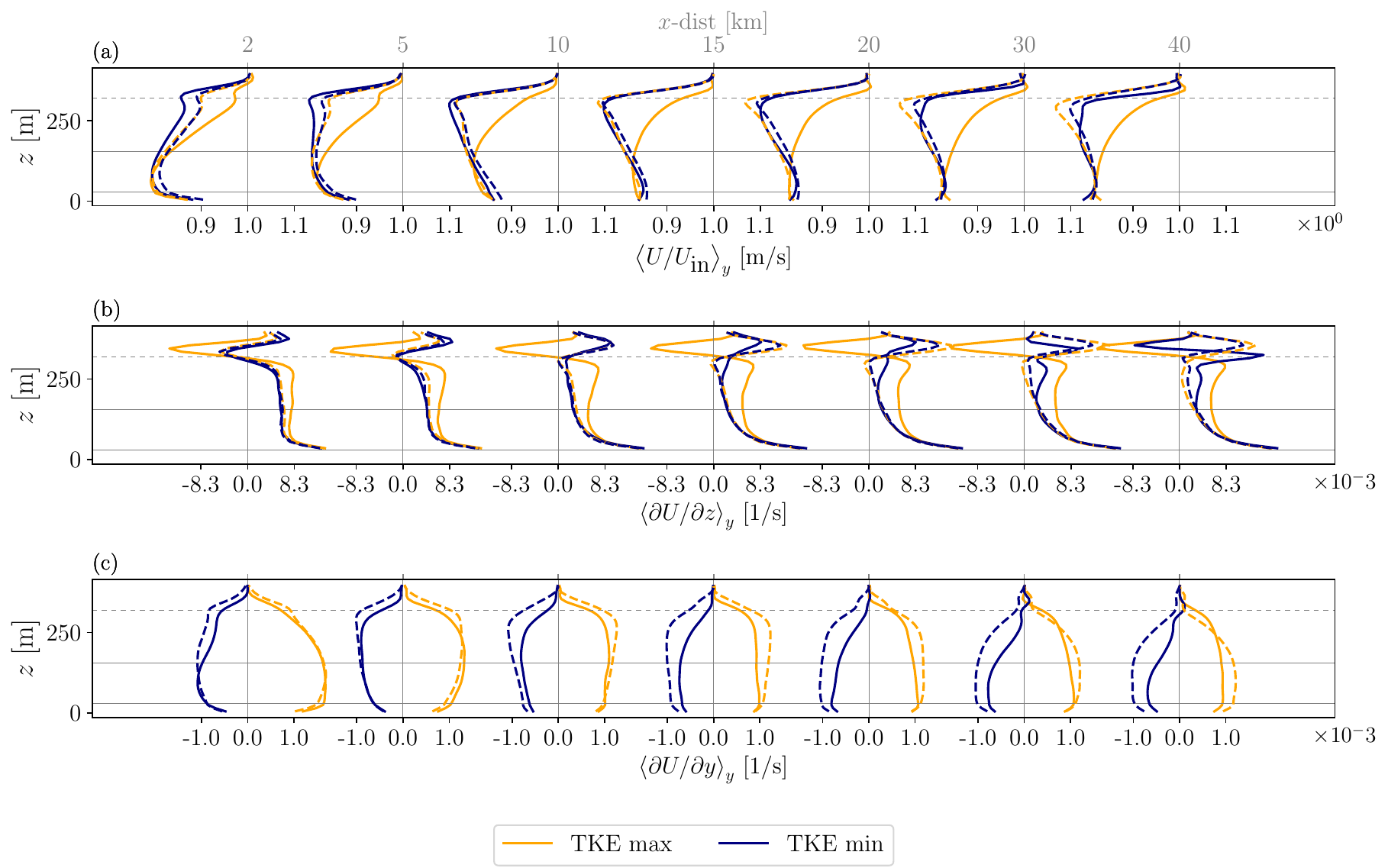}
    \caption{Vertical profiles of normalised velocity (a), vertical shear (b), horizontal shear (c) averaged at different $x-\textrm{dist}$ locations inside the left and right f.d. flanks of the wake for the NH (\rule[0.5ex]{.5cm}{0.5mm}) and the nE-nC (\hdashrule[0.5ex]{0.7cm}{0.5mm}{1mm}). Solid horizontal grey lines mark the wind turbines' bottom and top tip. Dashed horizontal gray lines the height of the ABL at the inflow. In panel (b), the vertical profiles are drawn only from 30\,m upwards as near the large value near the ground masked the relevant differences across the profiles.}
    \label{fig:shear_vertical_profiles}
\end{figure}

In Fig. \ref{fig:shear_vertical_profiles}, we deepen the analysis of the streamwise velocity behaviour and its crosswise and vertical gradients by illustrating the downstream development of vertical profiles at the TKE max and TKE min regions (as defined in Fig. \ref{fig:tke_production_bars} (a,b)) for the two cases NH and nE-nC. 
The streamwise velocity profile, shown in Fig. \ref{fig:shear_vertical_profiles} (a), is significantly larger in the top half of the ABL inside the TKE max region for the NH simulation. The same does not hold true for the southern side of the wake (TKE min) where the velocity profile of the NH simulation matches those of the nE-nC. The latter simulation shows minor to no differences across the two wake sides, as the lack of Coriolis does not induce an asymmetric behaviour in the wind farm wake.

A larger mean flow velocity in the top half of the ABL implies a larger vertical shear in the TKE max region of the NH simulation, as demonstrated by Fig. \ref{fig:shear_vertical_profiles} (b). 
When comparing the vertical shear profile across the two regions and two simulations considered, it becomes clear that the streak of higher TKE induced by veer is mostly produced in the region of the ABL above the turbines' hub height. This consideration is fundamental for understanding the origin of the streak. In fact, this region is characterized by a negative \Vin, thus $V$ at the left side of the wake in the NH simulation advects the free-stream with higher momentum towards the wake region. We thoroughly discussed this behaviour in \autoref{sec:streak_nature}.

Before continuing to investigate how veer induces a greater $\partial U /\partial x$ in the TKE max region, we consider the streamwise velocity gradient in the crosswise direction, $\partial U / \partial y$, included in \autoref{fig:shear_vertical_profiles} (c), as it is also responsible for a non-negligible TKE production at the wake sides.
In the TKE max region, this quantity is marginally smaller for the NH simulation than the nE-nC. We expect that this is the result of the larger momentum transport driven by the higher Reynolds stresses in this region.
Looking at the right side of the wake in the NH case, the horizontal velocity gradient is reduced at all heights of the ABL with respect to the nE-nC case. This phenomenon is not connected to faster diffusion but rather is based on the flow advection. As it is also discussed in \cite{Lanzilao_Meyers_2025}, we will not describe it here further.
Overall, the asymmetry we found in $P_{xy}$ in \autoref{fig:tke_production_bars} across the NH farm wake results from specific dynamics induced by veer at the side of the wake opposite to the one where the TKE streak exists. Therefore, we claim that $\partial U / \partial y$ is not involved in the formation of the TKE streak found in the numerical framework.

\subsubsection{Reason of larger velocity in the top half of the ABL in the TKE max region}\label{sec:streak_nature}
The main takeaway from the previous part is that veer induces the TKE streak by increasing the vertical shear $\partial U / \partial z$. This increase happens mostly in the top half of the ABL, and and is due to a larger $U$ aloft. The regions where the larger velocity is found, the left f.d. side of the wake for the NH and yE-nC cases, and the right one for the SH case, also correspond with a \Vin that is directed towards the wake region. Thus, in this section, we investigate the role of the crosswise flow in and around the farm wake and its implications in the TKE streak formation.
\begin{figure}
    \centering
    \includegraphics[width=\linewidth]{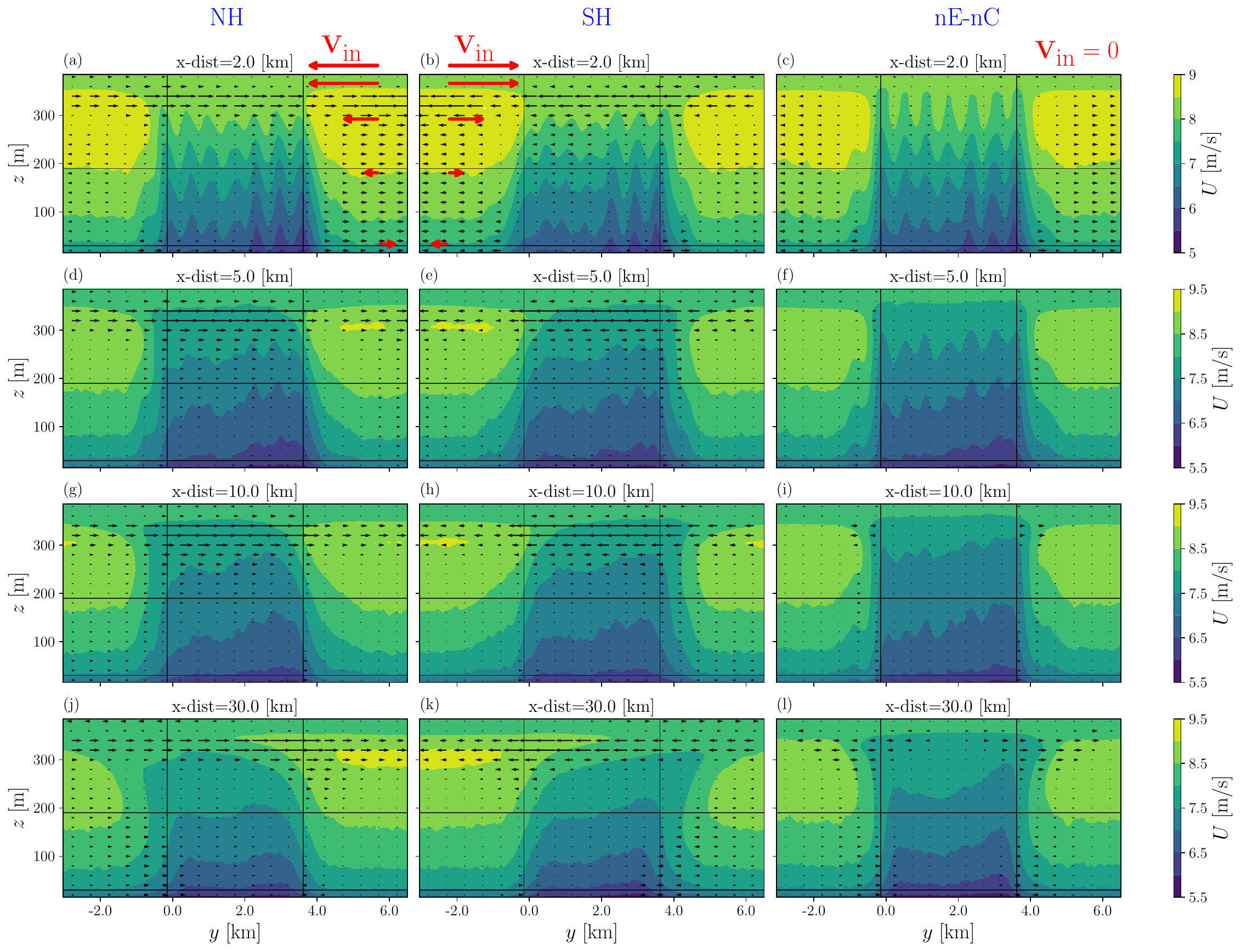}
    \caption{Streamwise velocity $U$ shown as coloured contours at several cross-stream planes downstream of the simulated wind farm for the NH (left panels a,d,g,j), SH (central panels b,e,h,k), and nE-nC (right panels c,f,i,l) simulations. The vector field of $V_\text{diff} = V - V_{\textrm{in}}$ is also provided with arrows to show regions of crosswise flow convergence at the left (right) side of the wake for the NH (SH) simulations. These patterns are absent for the nE-nC case. However, also in this simulation without veer, the farm wake promotes crosswise mean motion.}
    \label{fig:cross_planes_u}
\end{figure}

\autoref{fig:cross_planes_u} displays the streamwise velocity field together with a vector field of only the horizontal crosswise velocity perturbation, $V_{\textrm{diff}} = V - V_{\textrm{in}}$. In the figure, we show a schematic representation of \Vin with red arrows.
Inspecting the black arrows of panel (a) and (b) of \autoref{fig:cross_planes_u}, it is possible to observe that the right (left) f.d. side of the wake for the NH (SH) simulation experiences a slight positive (negative) perturbation of $V$ above the turbines' hub height, which causes convergence with the flow immediately outside the wake, whose $V$ is not particularly altered. The effect is exacerbated when moving downstream to 5\,km (d, e) and 10\,km (g,h). 
Such a convergence originates predominantly from a positive perturbation of the $V$ component within the wake region. In fact, the larger mixing promoted by the wake reduces the veer magnitude, inducing a positive perturbation above hub height where \Vin is negative. This effect is further amplified by the Coriolis force, which, in the wake, as a consequence of a deceleration of $U$, induces a positive acceleration in $V$. Furthermore, another effect that contributes to the $V$ perturbation inside the wake region is the slight upward displacement of the potential temperature inversion layer. Most of the wind direction change in the \Vin vertical profile happens, indeed, in the vicinity of this layer, where turbulence is greatly dampened by the buoyancy response. Therefore, when the inversion is displaced upwards, the lower fluid layers with less negative \Vin advect upwards, causing a large positive $V_{\textrm{diff}}$ just under the inversion layer. 

Beyond the 10\,km distance from the last turbine in the farm, another effect becomes evident. The wake shape starts to become more and more skewed. This behaviour must be attributed to the veer distribution in the incoming wind, schematically represented by the red arrows in Fig. \ref{fig:cross_planes_u}. In fact, above the turbines' hub height, the wake is progressively more advected towards the negative (positive) $y$ in the NH (SH) simulation. The result is that flow particles belonging to the free-stream next to the wake are advected over flow particles that are decelerated by the wake. This mechanism, together with the previously described convergence, causes the increase in the vertical shear identified in \autoref{sec:origin}.
\begin{figure}
    \centering
    \includegraphics[width=\linewidth]{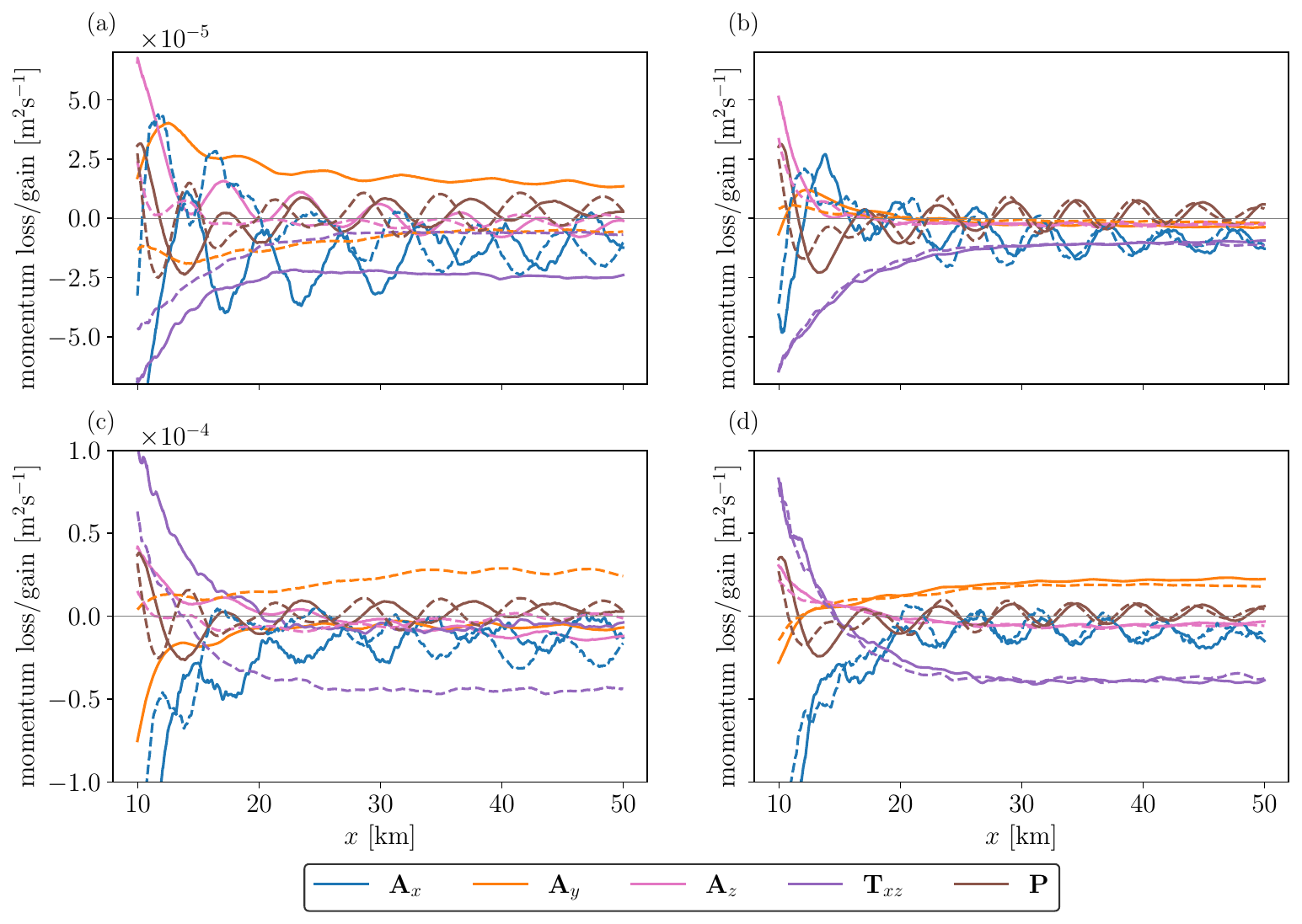}
    \caption{Most relevant terms in Eq. \ref{eq:wr_budgets} exhibiting differences across the NH (a, c) and nE-nC (b, d) cases that justify the horizontal convergence of momentum from which the TKE streaks originate. Each term is averaged in control volumes at the left (TKE max, \rule[0.5ex]{.5cm}{0.5mm}) and right (TKE min, \hdashrule[0.5ex]{0.7cm}{0.5mm}{1mm}) wake sides f.d., respectively, for two different height intervals: 150-300\,m (a, b) and 30-90\,m (c, d).}
    \label{fig:momentum_balance}
\end{figure}
To properly quantify how Coriolis effects induce an acceleration in $U$ in the top half of the ABL in the TKE max region, we compute the local momentum balance for $U$,
\begin{align}\label{eq:wr_budgets}
    -\underbrace{U\frac{\partial U}{\partial x}}_{\mathbf{A}_x}  
    -\underbrace{V\frac{\partial U}{\partial y}}_{\mathbf{A}_y} 
     -\underbrace{W\frac{\partial U}{\partial z}}_{\mathbf{A}_z} 
      -\underbrace{\frac{\partial}{\partial x_i} \left( \overline{u'_1 u'_i} + \overline{\tau^{\text{SGS}}_{1i}}\right)}_{\mathbf{T}_{xx_i}} 
      - \underbrace{\vphantom{\frac{\partial}{\partial x_i}}\varepsilon_{1jk}f_j (U_k-U_{g,k})}_{\mathbf{P_c}}
       -\underbrace{\frac{1}{\rho }\frac{\partial \pi*}{\partial x}}_{\mathbf{P}}
         = 0,
\end{align}
neglecting the buoyancy force because it has negligible effects in the control volumes later considered.
We average each term in \autoref{eq:wr_budgets} along the crosswise and vertical directions in the control volumes at the left (TKE max) and right (TKE min) wake sides for the NH and nE-nC simulations. As previously done, the control volumes are identified by the orange and blue outlines in Fig. \ref{fig:tke_production_bars} panels (a) and (b) for the two cases, respectively. In Fig. \ref{fig:momentum_balance} we further split the TKE max and TKE min control volumes in two, identifying one region below hub height, $z \in ]30,90[$\,m, and one in the top half of the boundary layer, $z \in ]150,300[$\,m. \\
The momentum balance shows that only in the upper part of the ABL there is a horizontal convergence ($\mathbf{A}_y$) that drives a larger streamwise velocity increase ($\mathbf{A}_x$) in the NH simulation (Fig. \ref{fig:momentum_balance} (a)). At the opposite side of the wake (TKE min), we find a slightly negative $\mathbf{A}_y$, signifying a divergence of momentum in the $y$ direction. This is associated with veer advecting the slow-moving flow in the farm wake core into the control volume we consider at the right wake edge. As a consequence, the wake recovery (negative $\mathbf{A}_x$) at this edge of the wake is hampered. Concerning the flow convergence in $\mathbf{A}_y$, the situation in the control volume under hub height is opposite to that in the control volume in the top of the ABL, as veer changes sign.\\
In the upper control volumes, the vertical turbulent transport is negative because turbulence drives momentum away and into the regions below (Fig. \ref{fig:momentum_balance} (c, d)). In the TKE max region of the NH case, this effect is exacerbated by the greater TKE, and the $\mathbf{T}_{xz}$ is more negative than in the TKE min region and at both wake edges of the nE-nC. Thus, the TKE streak promotes a larger wake recovery at lower heights, as demonstrated by the fact that $\mathbf{T}_{xz}$ in the TKE max of the NH case is the greatest in the lower control volumes. Here, the $\mathbf{T}_{xz}$ switches from positive to negative at a certain downstream distance from the farm. This is due to the fact that as the turbulence produced by the wind farm dissipates, the $\mathbf{T}_{xz}$ returns to behave as in the free stream, where it is negative as it must balance the positive pressure gradient of the geostrophic wind driving the flow (not shown in the plot). 

Analysing the trend of the pressure term helps to understand the wavy behaviour observed in all the advection terms, particularly in $\mathbf{A}_x$ and $\mathbf{A}_z$. These quantities are the most affected by the wind farm induced gravity wave at the capping inversion \citep{Allaerts_Meyers_2017, Lanzilao_Meyers_2024}. 
It is worth noting how the presence of veer affects also the propagation of such an interfacial wave. In fact, the oscillations between the left and right sides of the wake are not in phase for the NH simulations, but they are for the nE-nC simulation.\\

A final consideration is that the mechanism through which the TKE streak is formed is not strictly dependent on the Coriolis force, nor the type of obstacle creating the wake. 
Thus, according to our results, the wake of any obstacle invested by a flow with veer should present a TKE streak at the side where the $V$ velocity component in the undisturbed veer profile points mostly towards the wake region. The additional presence of shear can further enhance the effect and add the condition for the TKE streak appearance that $V$ should point towards the wake in the top part of the shear profile, where $U$ is larger.
However, a large wind farm operating in a marine ABL is an exemplary configuration in which TKE streaks appear. In the following, we focus on whether the presence of such streaks can positively affect the recovery of the wind farm wake in a shallow ABL.

\subsection{Implications on wake recovery}\label{sec:wr}

\begin{figure}
    \centering
    (a)
    \includegraphics[width=0.45\linewidth]{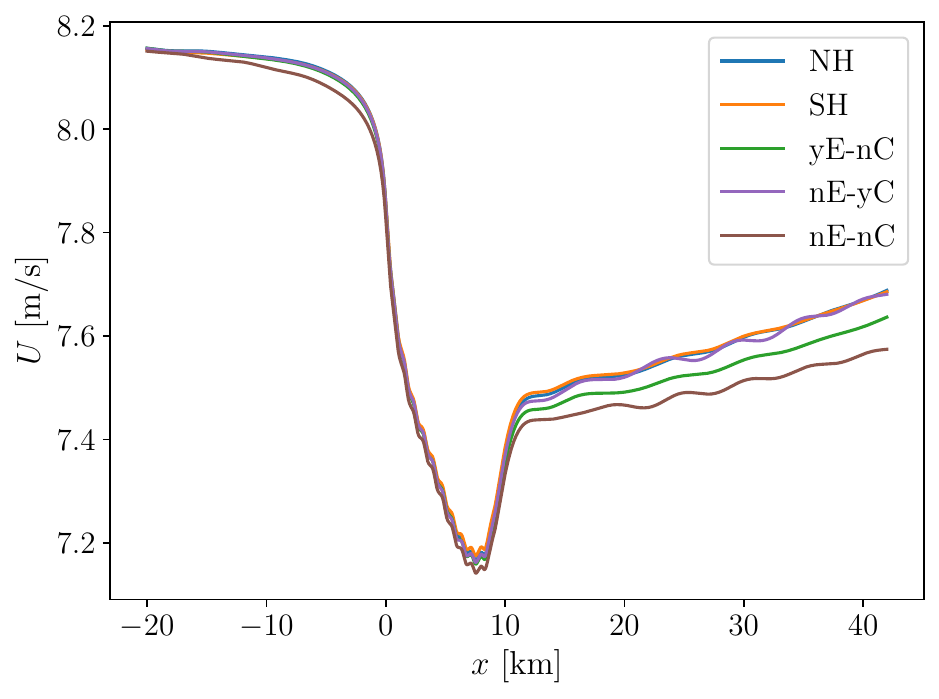}
    (b)
    \includegraphics[width=0.45\linewidth]{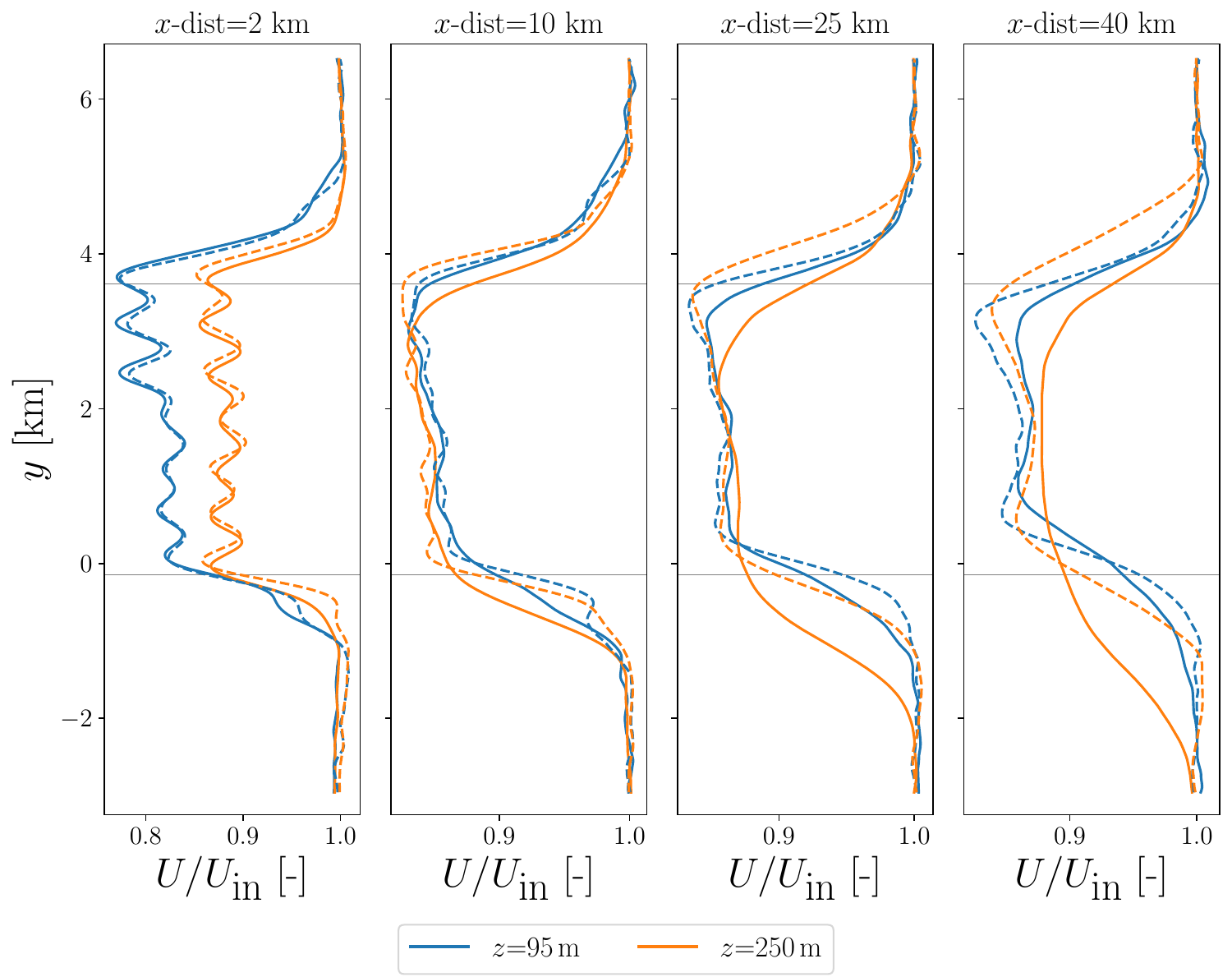}
    \caption{Mean streamwise velocity component $U$ averaged in $z,y$ in a control volume around the wind farm for all the main run simulations performed (NH simulation is mostly completely overlapped by SH) (a). Mean streamwise velocity component $U$ plotted along the crosswise direction $y$ for different distances downstream the last turbine of the wind farm (b) for two different heights (line colours) and for the NH (\rule[0.5ex]{.5cm}{0.5mm}) and the nE-nC (\hdashrule[0.5ex]{0.5cm}{0.5mm}{1mm}). Horizontal grey lines mark the wind farm extent along $y$.}
    \label{fig:ws_profiles}
\end{figure}

As a final analysis, we investigate how the wake of the wind farm recovers in the presence of the TKE streak induced by veer.\\
In Fig. \ref{fig:ws_profiles} (a), we illustrate the velocity averaged in a control volume around the wind farm, $y \in [-2, 5.8]$\,km and $z\in [10, 300]$, for all the main run simulations performed. The choice of such a wide control volume around the wind farm is taken to prevent veer from advecting the wake outside the control volume.

The streamwise velocity profiles are very similar upstream and inside the wind farm across all the simulations, except for the nE-nC case, which exhibits more global blockage upstream of the wind farm. In the farm wake, the differences across the profiles are more evident. Again, the simulation with neither veer nor the wake-deflecting Coriolis force differentiates the most from the others. In this case, the wake recovery seems significantly lower. The fast near-wake recovery driven by turbulence stops earlier, and the far-wake recovery, which is linear in our simulations, is lower. These differences are not motivated by different power extraction at the wind turbines across the cases. In fact, we find that the differences in the total mean wind farm power are less than 1\% across the cases.
Therefore, the faster wake recovery in the simulations with veer must be a property of the farm wake. Interestingly, the lack of the w.d. Coriolis force seems to penalize wake recovery more than the lack of veer, as shown by the fact that the nE-yC case shows overall higher velocities in the wake than the yE-nC case.

We also investigate several crosswise wake profiles for the NH and nE-nC cases in Fig. \ref{fig:ws_profiles} (b). At hub height (light blue curves), in the NH simulation, $V(z=z_\text{hub})= 0$, as shown by \autoref{fig:inflow} (b). Hence, the horizontal velocity vector is the same for the NH and nE-nC simulations. However, the fact that the wake of the wind farm in the NH case propagates in a similar direction to the nE-nC simulation is justified by the fact that the entrainment of veer from the upper layer compensates for the deflection induced by the Coriolis force arising from the flow deceleration in the wake.\\
The evolution of the horizontal wake profiles suggests that the wind farm wake in the NH case has a superior rate of recovery. In particular, the horizontal gradients at the wake edges appear smoother in the NH simulation.
At the left side, the higher vertical turbulent transport is responsible for stronger recovery (Fig. \ref{fig:momentum_balance} (c)). At the right side, as observed by \citet{Lanzilao_Meyers_2025} (see Fig. 11 in their paper), the velocity gradient from the wake to the free stream is lower, with a transition to the free stream smeared out over a greater $y$ distance. This effect is exacerbated at higher heights, $z$=250\,m.\\
Considering now the velocity in the wake at 250\,m. At only 2\,km downstream of the wind farm, the wake deficit is lower than that at hub height. At this stage, the wake is still propagating upward. Once the maximum wake deficit reaches the capping inversion (somewhere between $x$-dist = $[2, 10]$\,km), the wake is rather homogeneous with height, as demonstrated by the curves overlap at 10\,km downstream (Fig. \ref{fig:ws_profiles} (b)).\\
While moving further downstream, the NH wake appears more and more skewed. Mostly, this behaviour is justified by a South-West propagation direction imposed in the top part of the ABL by the background flow veer. The lack of Coriolis effects, represented in the nE-nC simulation, implies that the direction in which the wake propagates is the same at different heights.\\
However, in the nE-nC simulation the wake at $z$=250\,m shows smoother crosswise gradients to the free stream. This behaviour can be interpreted by considering the secondary circulation observed in \autoref{fig:cross_planes_u} (i, l) and \ref{fig:streamlines} (i, l), that in the top half of the boundary layer advect flow along the $y$ direction from the wake to the free stream. These secondary flow structures should be further investigated as they may influence the farm wake expansion in the horizontal plane.
Overall, our results show that the TKE streak present on the left side of the wake in the NH simulation influences the recovery of the wake. In fact, the crosswise velocity gradient at hub height in the TKE max region is lower in the NH simulation than in the nE-nC simulation, and the velocity deficit is also lower. This indicates that the higher vertical momentum entrainment in the lower part of the ABL, promoted by the TKE streak (\autoref{fig:momentum_balance} (c)), locally affects wake recovery.
However, the presence of the Coriolis force has a significantly larger impact on the overall recovery of the wake than the TKE streak itself. As demonstrated by the fact that, even at the right side of the wake (TKE min), the NH simulation shows a smoother shear layer and a higher velocity than the nE-nC simulation.

\section{Discussion of streaks in LES and streaks observed in the German Bight}\label{sec:discussion}

The results presented so far demonstrate that the TKE streak in LES is an implication of the wake deflection imposed by veer and to a minor extent the Coriolis force deflecting the wake. Therefore, as long as the boundary layer is shallow, thus characterized by large change of wind direction with height, and spatially homogeneous we can expect the TKE streak to appear in the wake of real wind farms or other similarly sized objects. 
The purpose of this section is to discuss if the streak observed by the airborne measurements contained in \citet{Platis.2018} and the streaks in the SAR scenes of \autoref{fig:SAR_panels} are connected with the LES TKE streak.

As already mentioned in the introduction (\S \ref{sec:intro}), \citet{Platis.2018} found through airborne measurements a persistent TKE streak at the western side (left f.d.) of the the wake of a wind farm cluster in the German Bight operating in a southerly wind. 
Their observation agrees well with the LES findings. Firstly, the streak is located in a very similar position as what is suggested by LES. Secondly, the localized increase in TKE found inside the streak in the airborne measurements is of the same order of magnitude as the one we found with LES. Although, quantitatively, the increase in TKE within the streak is larger in the measurement than in the LES. 
At this point, it is worth reminding that a large horizontal background flow gradient was present at the time of the measurement flight. 
According to \autoref{fig:tke_prod_components}, the crosswise gradient of the streamwise velocity plays a secondary role in the overall TKE production in the wake of a wind farm, even at the wake edges, where this gradient is maximum. 
Thus, we expect that the background flow gradient is not the only factor producing the TKE streak observed by the airborne measurement.
That instead, it could be formed by a combination of the effect of veer, through the same mechanism we detailed in \autoref{sec:streak_nature}, and the TKE production from the background flow velocity crosswise gradient. \\
Overall, we interpret the observation presented in \cite{Platis.2018} as supportive of the LES results presented in this study.

We now focus on comparing LES to SAR. It is evident that several striking similarities exist between the higher NRCS streaks shown by SAR (\autoref{fig:SAR_panels}) and the streak of higher TKE found in LES. 
The fact that, regardless the wind direction, the same irregular wind farm layout consistently produces NRCS streaks always on the left f.d. side of the wake in the northern hemisphere (\autoref{fig:SAR_panels}) is difficult to justify without taking into consideration the Coriolis effect. 
However, this is not sufficient to demonstrate that the same physical explanation of the LES TKE streak provided in \autoref{sec:streak_nature} holds also for the NRCS streak observed in SAR.
In fact, SAR is not measuring TKE directly, nor it provides a measurement of the flow at hub-height. SAR is expected to measure the friction velocity $u^*$. Thus to prove a connection between the streaks observed in SAR to the one in LES the NH simulation should show an increased turbulent shear stress $\tau \propto u^*$ at the left side of the wake. Unfortunately, \autoref{fig:streak_identification} shows that the turbulent shear stress behaves similarly to the TKE at hub height but not near the ground, where it is a proxy for $u^*$.

At this point, an important fact we need to take into account is that a typical simulation with PALM cannot describe accurately the friction velocity or the shear stress near the ground \citep{Maronga2020b}. To understand why, it helps recalling how the MOST boundary condition is applied to the momentum equation.
According to MOST, the friction velocity can be derived integrating the following equation from the surface roughness height $z_0$ to the first grid point height $z_1$:
\begin{equation}
    \frac{\partial u_{\textrm{h}}}{\partial z} = \frac{u^*}{\kappa z}\Phi_m\left( \frac{z}{L}\right),
\end{equation}
where L is the Monin-Obukhov length, $\kappa =0.4$ the von Karman constant, and $u_{\textrm{h}}$ the horizontal velocity at the ground, $(u^2 + v^2)^{\frac{1}{2}}$, and $\Phi_m$ a stability dependent parameter that we will further omit as potential temperature changes are negligible inside the boundary layer, hence $\Phi_m(z/L) = 1$. The integration yields:
\begin{equation}
    u^* = \frac{\kappa\, u_{\textrm{h}}}{\ln\left(\frac{z_1}{z_0}\right)}
\end{equation}
However, $u^*$ is also connected to the vertical momentum flux as per \autoref{eq:us}. This gains a boundary condition for the diffusive term of the momentum equation at the wall of the type
\begin{equation}\label{eq:wall_model}
\begin{cases}
     \tau_{xz} = u^{*^2} \cos{\alpha} =  \frac{\kappa^{2}}{\ln\left(\frac{ 2 z_1}{z_0}\right)\cos{\alpha}} u^2_{z=z_1}\\
     \tau_{yz} = u^{*^2} \sin{\alpha} = \frac{\kappa^{2}}{\ln\left(\frac{ 2 z_1}{z_0}\right)\sin{\alpha}} v^2_{z=z_1}  
\end{cases}
\end{equation}
with $\alpha$ being the wind direction at the first grid point height $z_1$.
Unfortunately, the above relations are compromised by the fact that the resolved scale near the wall contains very little energy. In fact, turbulence scales do scale with distance from the wall \citep{pope2001}, and the constant grid spacing in our LES means that a large part of the turbulence close to the wall is unresolved. 
In boundary layers wall modelled LES, this is a common issue, as pointed out by \citet{Kawai2012, Maronga2020b}. Both the authors reported that, as the resolved velocity lacks a large part of the turbulent contribution, the friction velocity, as well as the shear stress near the wall, is underestimated. The result is that the streamwise velocity exceeds what is calculated by MOST above the first grid point of the computational domain.

In our particular case, in \autoref{fig:streak_identification} (f) we show the time averaged mean shear at 10\,m. This value is strongly affected by the boundary condition $\tau_{xz}$ and $\tau_{yz}$.  A time average of Eq. \ref{eq:wall_model} gains
\begin{equation}
    \begin{cases}
    \overline{\tau_{xz}} =  \gamma \overline{uu}  = \gamma \left(UU + \overline{u'u'} \right) \quad\quad\quad \textrm{with}\quad\quad\quad\gamma=\frac{\kappa^{2}}{\ln\left(\frac{2 z_1}{z_0}\right)\cos{\alpha}} \\
    \overline{\tau_{yz}} =  \beta \overline{vv}  = \beta \left(VV + \overline{v'v'} \right) \quad\quad\quad \textrm{with}\quad\quad\quad\beta=\frac{\kappa^{2}}{\ln\left(\frac{2 z_1}{z_0}\right)\sin{\alpha}}.
    \end{cases}
\end{equation}
This means that the momentum flux at the wall, e.g. the shear stress $\tau$ set by the wall parametrization, is mostly dependent on the mean kinetic energy of the flow $U_iU_i$ as $U_iU_i >> \overline{u'_iu'_i}$. This steers the calculation of the instantaneous velocity field towards a solution where the shear stress at the wall is mostly proportional to the mean flow field. As in our simulations, no streak exists in the mean velocity, we speculate that the TKE streak is not translating into the $\overline{\tau}$ near the wall due to the wall model.
 
To mitigate the effect of the wall model on the solution, \citet{Kawai2012} and \citet{Maronga2020b} proposed that friction velocity could be calculated from \autoref{eq:wall_model} by using the resolved velocity from a grid point located further away from the wall, but still inside the surface layer ($\sim$10\% of ABL height). This choice is motivated by the fact that the shear stress in the surface layer is rather constant. Thus, it is possible to derive $\tau_w$ from the value of $\tau$ at a higher height where the wall model affects less the solution and turbulence is better resolved. \citet{Maronga2020b} found that the two previous hypothesis are generally satisfied at the seventh grid cell above the domain bottom surface. 
Unfortunately, in our LES set up, the cell centre of the seventh grid element is located at 65\,m and outside the surface layer (estimated to be $\sim$35\,m according to the famous 10\% of the boundary layer height criterion). 

With the set-up used in this study we are not able to verify if the method described above could demonstrate that a proportionality of TKE and $\overline{\tau}$ exists also near the ground.
Therefore, we conducted an additional simulation of the NH case where the vertical grid spacing within the boundary layer was reduced from 10\,m to 4\,m.  With this finer resolution, the surface layer of the simulated ABL is discretized by  seven grid points. Thus, although the near wall region is still largely under-resolved, it exists a grid point within the surface layer where to compute a $\overline{\tau} \propto \overline{\tau_w}$, that is expected to be no longer affected by the wall model (according to \citet{Maronga2020b}). However, even at this grid point, the shear stress $\overline{\tau}$ does not show a clear streak at the left side of the wake (not shown here). Thus, we could not prove a correlation between TKE and $\overline{\tau}$ near the ground with this higher resolution. This means that according to our LES results, the higher TKE near the ground should not correlate to an increase in the $\tau_w$, and hence it should be invisible to SAR.\\
Given the several similarities existing between the TKE and NRCS streaks, under the hypothesis that the wall model is not artificially decoupling TKE and $\tau_w$, still to us it remains an open question if the SAR NRCS is only sensitive to the friction velocity, and if it is in no measure affected by the TKE. This topic stretches far beyond the scope of the current paper. Thus, next, we focus on discussing several other complications in a direct comparison of the LES results with SAR that hinder demonstrating a connection between the TKE streak and the NRCS streaks.

First, we address the validity of comparing temporally averaged quantities from LES with the instantaneous observations provided by SAR. A major source of uncertainty in field measurements at large offshore wind farms arises from the vast spatial extent that the measurements have to cover, which prevents the acquisition of densely sampled data in both time and space. While SAR offers an exceptional capability to capture large areas at high spatial resolution, it is inherently limited by relatively sparse temporal sampling, with revisits typically every day or two, depending on the number of available satellites.\\
Instantaneous fields of a quasi-steady-state LES often do not offer much insight. Therefore, we have not presented them in the body of the paper. However, the instantaneous LES fields of the fluctuating velocity components ($u'_i$) at hub-height also reveal the characteristic streak at the left wake edge for the NH case. \\
However, one seemingly necessary condition for the formation of persistent streaks is the lack of significant spatial and temporal variability in the atmospheric flow, as was the case for several of the SAR observations with visible streaks that we examined.
Under this condition, the physics clarified by LES, mainly involving the mean flow, should still be visible in the instantaneous observation provided by SAR.

Another important point not discussed yet is that the simulations we performed do not account for the physics of the sea surface that SAR measures. In fact, we do not simulate the waves at the air/water interface caused by the wind. Nor do we apply models that describe the changes in surface roughness with different near-sea-surface wind speeds \citep[e.g][]{Charnock1955}.\\
Simulating waves is extremely challenging, and it is commonly avoided in highly resolved LES simulations of offshore wind farms.
Using instead a model like the one of \citet{Charnock1955} could be troublesome because it is based on the hypothesis of quasi-steady waves, i.e. the wave height, and hence surface roughness, is achieved after a long fetch. It is unclear how this assumption can be met in the wind farm wake, where the velocity increases continuously with increasing distance from the wind farm. Although we do not expect that simulating waves would change our results fundamentally, we cannot rule out that it could be the key to interpreting the SAR measurements.

Finally, we discuss the apparent discrepancy in the fact that our simulations are conducted in a CNBL, but according to \autoref{fig:SAR_parameter-map} NRCS streaks appear more often with stable boundary layers. 
Stable boundary layers tend to have less turbulence than neutral or unstable ABL. As a result, an SBL has a larger veer because of the reduced vertical mixing.  
Thus, the mechanism described in \autoref{sec:streak_nature} should produce even stronger TKE streaks in a stable boundary layer where veer is more intense.
The underlying assumption here is that the buoyant forces in the ABL would not significantly change the way the TKE streak is formed in the TKE max region.
So, if on the one hand, we expect that in a SBL the TKE streak could become even more pronounced, it is an open question if it could better correlate with $\tau_w$ in such ABL type.
Unfortunately, the wall model limitations presented above are expected to worsen in SBL as the amount of resolved turbulent fluctuations is even lower with the same grid resolution.

The final exercise we conduct to conclude this section is the analysis of the counter argument that the streaks of higher NRCS observed by SAR are due to streaks of increased velocity rather than TKE. 
Speed-ups at the edge of the wake have been observed through lidar and turbine operational data coming from the Supervisory Control and Data Acquisition (SCADA) system in \cite{Schneemann2020.b} and \cite{Nygaard.2016}, respectively. Unfortunately, in both studies, only the left f.d. side of the wake is characterised, and it is not possible to conclude whether the acceleration happens only at one or both sides of the wake. Thus, we will not consider this observation further.
Also \citet{Marwa2025} found a speed-up at the sides of the wake for an idealized wind farm placed in the northern hemisphere. Under extremely shallow SBLs, the speed-up at the left side f.d. of the wake was greater. The speed-ups were justified by the fact that the inversion layer of their SBL is so close to the turbine tips that it impedes the flow displaced by the turbines from escaping above the wind farm. In such a case, it is possible that the same physics described by \citet{Grisogno2004} could play a role, despite the moderate crosswise extent of the farm.\\
Other previous numerical studies \citep[c.f.][]{Hasager.2024}
compared the wind velocity field at 10\,m calculated by the Weather Research \& Forecast Model (WRF) in the wake of offshore wind farms and the co-located velocity fields derived from SAR measurements, finding a good match on some peculiar flow feature. However, such flow feature do not stretch as long as the streaks displayed in Fig. \ref{fig:SAR_panels}, and they could be justified by the enhanced vertical momentum transport in the near wind farm wake. This effect was also discussed by \citet{Djath.2018}, and they already proposed a simple parametric model to describe the enhanced downward momentum flux suggested by SAR imagery, typically limited to 10\,km downstream of the current offshore wind parks in stable conditions.\\. 
In other WRF results, like the one presented by \citet{Siedersleben.2018}, meant to replicate the already discussed flight measurements presented by \cite{Platis.2018}, a streak of higher velocity appears on the left side of the wind farm cluster wake with South-Westerly winds (see p.412). The same velocity streak, however, is not found in the airborne measurements.
It is to be determined how the resolved higher TKE in the streak discussed throughout the current manuscript would appear in numerical models where turbulence is not resolved. For example, in the RANS simulations of \citet{vanderLaan2025damping}, streaks in the velocity field at the left side of a wind farm wake were deemed numerical artefacts, as they exhibit unphysical behaviour. 
Furthermore, \cite{Hasager.2024} showed that the near-surface wind speed-up downstream of offshore wind farms matched between SAR and the WRF mesoscale simulations only when turbine-induced TKE is included. This supports the view that SAR signatures in wakes are strongly influenced by turbulence processes directly, or through a modification of the flow velocity at the sea-surface. 
Overall, we believe that in the particular case of the wake of a wind farm in a shallow boundary layer, it may be misleading to interpret any flow feature observed by NRCS uniquely as a velocity change at the sea-surface.

To summarise the considerations we have discussed in this section, a connection between the streaks seen in SAR and in LES is hard to demonstrate. The LES set-up we use is intrinsically under-resolved near the wall, and the wall model affects its results. While we do not prove formally that the TKE streak increases the friction velocity at the ground, we prove that the wall model is responsible for decoupling these quantities.
Furthermore, we present several other points that could contribute to not observing a direct connection between TKE and NRCS streaks. 
Overall, given the several similarities across the two, we believe that a connection exists. However, a further study could focus on demonstrating the connection in a more systematic manner. For example, by employing other remote sensing techniques, e.g. lidar, to sample the flow near the ground.\\
Besides the several open questions in connecting SAR and LES, we found a qualitative agreement between our LES results and the airborne measurements shown by \citet{Platis.2018}.

\section{Summary and Conclusions}

Within this study, we investigate with LES how the Coriolis effect introduces an asymmetric behaviour in the turbulence kinetic energy field of offshore wind farm wakes. The asymmetry is represented by a streak, located only at the left edge of the farm wake along the flow direction (f.d.) for the simulation of the wind farm in the northern hemisphere, where the TKE is larger than in both the free stream and the remaining wake region.

To investigate if and how Coriolis causes such a streak, we performed simulations of the same real wind farm in the northern and southern hemisphere, and also in three additional fictitious boundary layers where Coriolis effects, i.e. veer and the wake-deflecting Coriolis force resulting from mean flow changes, are switched off individually or all together.\\
With this set of simulations, we are able to identify the presence of veer in the incoming wind as the main contributor to the appearance of a persistent TKE streak. We found that the side at which the streak appears, left f.d. edge of the wind farm wake in the northern hemisphere, switches to the right f.d. in the southern hemisphere. The contribution of the Coriolis force resulting from mean velocity changes (induced by the turbines) in the TKE streak formation is only minor. Without veer, the TKE streak fades before the outlet of the computational domain. When the Coriolis effects are not present, the two sides of the wake exhibit a very similar behaviour, and symmetry is restored. 

Our results suggest that the impact of asymmetries in the wind farm layout on the formation of the TKE streak is not dominant. In fact, although the wind farm simulated has a non-symmetric layout in the flow direction, the TKE streak switches wake edge side when the farm is translated in the southern hemisphere, or even disappears when Coriolis effects are removed.
Further systematic studies are necessary to conclude how the farm layout affects the shape and intensity of the streak. Furthermore, the same wind farm should be simulated under different wind directions, e.g. south-westerly or north-westerly winds. This will allow to understand the behaviour of the streak when the layout of the wind farm promotes smoother velocity gradients at the wake edges.

Our results also allow us to identify the mechanism that forms the TKE streak. We found an increase in turbulence production driven by a larger vertical shear of the streamwise velocity ($\partial U / \partial z$) only at the wake side of the TKE streak. The larger shear is found at heights greater than the turbine's hub height, and it is the result of two mechanisms related to veer. The primary effect is that veer skews the wake region by affecting its propagation direction with height; this translates into the advection of unwaked flow in the top half of the boundary layer over the wake region in the bottom half. The secondary effect is that veer is reduced in the wake due to higher mixing, which causes a further convergence of momentum in the horizontal plane in the top half of the ABL, resulting in a streamwise speed-up and further vertical motion. Both these effects are sustained as long as the velocity deficit in the wake exists. Therefore, the streak is very stable.

An important point not yet discussed is that, despite being specifically impactful in wind energy, the physical phenomenon described here is a fundamental property of flows with a veer distribution. Although the Coriolis force in atmospheric flows is known to induce veer, similar veered flows could be created in other ways. Therefore, our results could be generalized to other domains of fluid mechanics. In particular, in applications where more mixing is required in a specific region of the flow, we suggest that, behind an obstacle subjected to veered inflow, more TKE production can be induced at the side of the wake where veer is converging.

Another finding of this study is that a secondary flow arises at the wind farm wake edges. In particular, we find a vortex core that is fully aligned with the TKE streak in both the NH and SH simulations. By comparison with the case without Coriolis effects (nE-nC), we understand that the secondary flow does not play a significant role in inducing the presence of the streak itself. In fact, we found that, in the nE-nC case, two vortex cores exist at both wake edges, from approximately 10 km downstream of the wind farm to the end of the simulated domain, and they do not induce any TKE streak. We argue that these features are Prandtl secondary flows of the second kind, i.e. driven by gradients in the covariance of the Reynolds stresses ($\varepsilon_{1jk}\partial_{x_j} \overline{u'_ju'_k}$) along the direction of the vorticity vector of the circulation \citep{Anderson_2015}. Future research should aim to verify our hypothesis and to better understand the impact of these features on farm wake expansion in the horizontal plane.

Finally, we consider the wake recovery for the wind farm simulated in the several cases studied.
Our results suggest that the Coriolis force has a visible impact on the wake recovery. In our results, the lack of both veer and the Coriolis force deflecting the wake induces the slowest wind farm wake recovery. Interestingly, the Coriolis force deflecting the wake seems to be more beneficial to recovery than the presence of veer. However, this behaviour is not well understood, and the two effects do not combine linearly in determining the recovery of a wind farm wake.\\
The boost in recovery offered by the TKE streak is also noticeable. In fact, we find that in the NH case, the wake horizontal gradient ($\partial U / \partial y$) towards the free stream at the left side f.d. is smoother than in the nE-nC case. Although the effect is quite minor, we expect that the power yielded by a second wind farm situated downstream is positively impacted by the presence of the streak.
Furthermore, the larger localized TKE should promote faster turbine wake recovery inside the second wind farm, again enhancing yield.
At the same time, the turbines in a hypothetical downstream farm yielding a larger power will also encounter significantly higher fatigue loads due to TKE levels up to 100\% larger than other turbines affected by the same farm wake but outside the streak region. Overall, a simulation with two wind farms is necessary to properly quantify the impact of the TKE streak on scenarios where several wind farms are installed next to each other, such as in the German Bight.\\

\noindent
\textbf{Acknowledgements.} \\
    The Authors would like to thank M. D\"orenkämper for suggestions at the stage of the work inception.
    Paul van der Laan and Thomas Messmer for fruitful discussion and revision of the manuscript, and Marcel Bock for further revision.
    The authors thanks Chat AI (Meta LLaMa) for supporting with matplotlib in the plots preparation and the final proofreading of the text.\\

\noindent
\textbf{Funding} \\
   This work is funded by the Federal Ministry for Economics Affairs and Climate Action according to a resolution by the German Federal Parliament (project \emph{C\textsuperscript{2}-wakes}. The authors gratefully acknowledge the computing time granted by the Resource Allocation Board and provided on the supercomputer Emmy/Grete at NHR-Nord@G\"ottingen as part of the NHR infrastructure. The calculations for this research were conducted with computing resources under the project \emph{nik00084}.\\

\noindent
\textbf{Author contributions} 
GC prepared the first draft of the paper. GS conceptualized the research. 
GC and GS conducted the body of the research. 
BS and JS provided SAR data and the SAR data analysis. 
GS and JP supervised the research. 
All the authors supported the revision and completion of the text body.
GC and GS contributed to securing the funding for the C$^2$-Wakes project 
and the computing resources at NHR-NORD@G\"ottingen.\\

\noindent
\textbf{Declaration of interests}
    The authors report no conflict of interest.\\


%

\appendix

\section{Decomposition of TKE production contributions $P_{ij}$}\label{sec:appendix1}
\begin{figure}
    \centering
    \includegraphics[width=\linewidth]{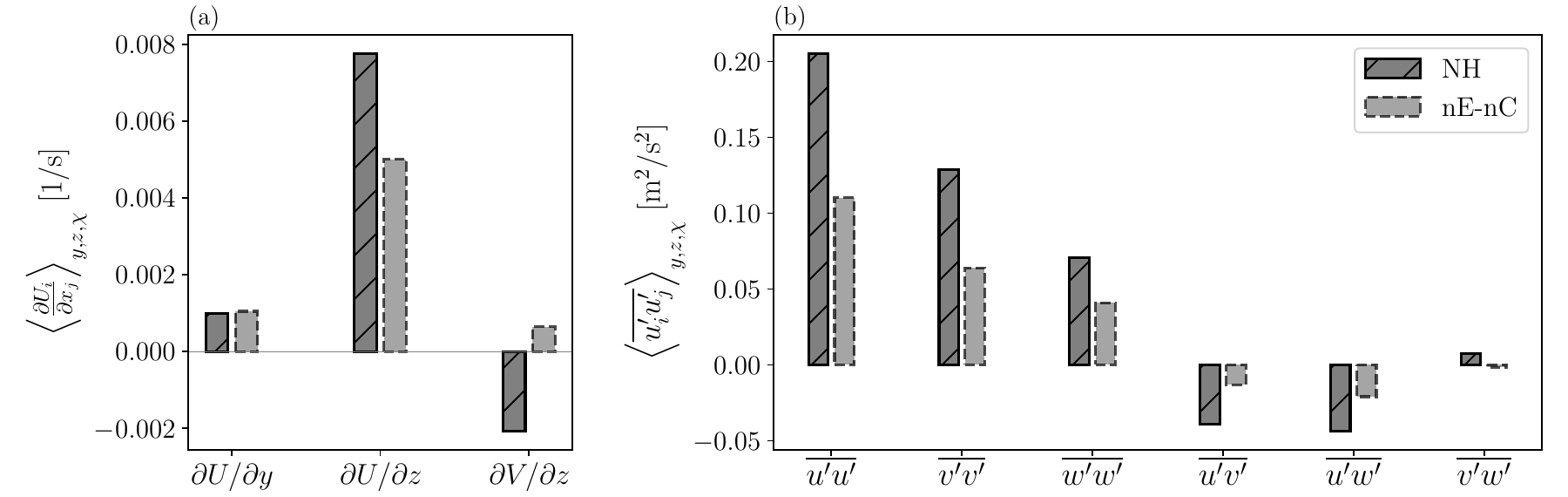}\\
    \caption{Components of the mean deformation tensor (a) and Reynolds shear stress tensor (b) averaged throughout the left side of the wake (TKE max) from 20 to 300\,m in the vertical direction for the NH and nE-nC simulation}
    \label{fig:tke_prod_components}
\end{figure}

In this appendix, we split the TKE production components $P_{ij}$ into the mean deformation matrix (Fig. \ref{fig:tke_prod_components} (a)) and Reynolds stresses  (Fig. \ref{fig:tke_prod_components} (b)), for only the relevant terms in \autoref{fig:tke_production_bars}. 
This helps better isolating the role of the larger shear, identified as the driver of the larger TKE streak in the paper body.\\
\autoref{fig:tke_prod_components} reveals that the larger TKE production at the left f.d. wake side in the NH case is induced by the significantly larger vertical shear ($\partial U / \partial z$). 

The crosswise gradient of the streamwise velocity is comparable across the simulations, but quite interestingly, the simulation without w.d. Coriolis force and without veer still exhibits an average non-zero vertical gradient of $V$. This is the result of a vortex core settling at the edge of the wake shown in Appendix \ref{app:secondary_flow}.

\section{Secondary flow of the second kind in the farm wake}\label{app:secondary_flow}
Within this appendix, we justify some flow features found in the results in \autoref{sec:results}, specifically in \autoref{fig:cross_planes_u}, that, despite not being connected to the formation of the TKE streak, are in our opinion connected to the turbulence properties of the wind farm wake.

As shown by Fig. \ref{fig:TKE_coriolis}, in all the main run simulations performed, the TKE in the farm wake is lower than in the free stream after a few km ($\sim$ 8\,km) downstream of the wind farm. This behaviour stems directly from the larger vertical mixing found in the near farm wake promoted by the turbines' added TKE. The larger vertical mixing is responsible for a reduction of the horizontal velocity gradient with height \citep{Lanzilao_Meyers_2025}, hence a lower TKE production must be expected within the wake region, as shown by Fig. \ref{fig:tke_budgets}. \citet{Anderson_2015, Wangsawijaya_2020} show that horizontal heterogeneity in the TKE of a boundary layer causes Prandtl secondary flow of the second kind.
\citet{Wangsawijaya_2020} shows that above the interface between smooth and rough surface stripes aligned with the mean velocity vector, a vortex forms as a result of turbulent transport unbalances across the regions with high TKE over the rough stripes and low TKE over the smooth stripes.

As clarified by \citet{Anderson_2015}, the main physical mechanism is the increase in the vorticity field in the $x$ direction, $\omega_x$, by the Reynolds stresses covariance:

    $$\omega_x \sim \varepsilon_{1jk}\frac{\partial \overline{u'_ju'_k}}{\partial x_j}.$$

\begin{figure}
    \centering
    \includegraphics[width=0.9\linewidth]{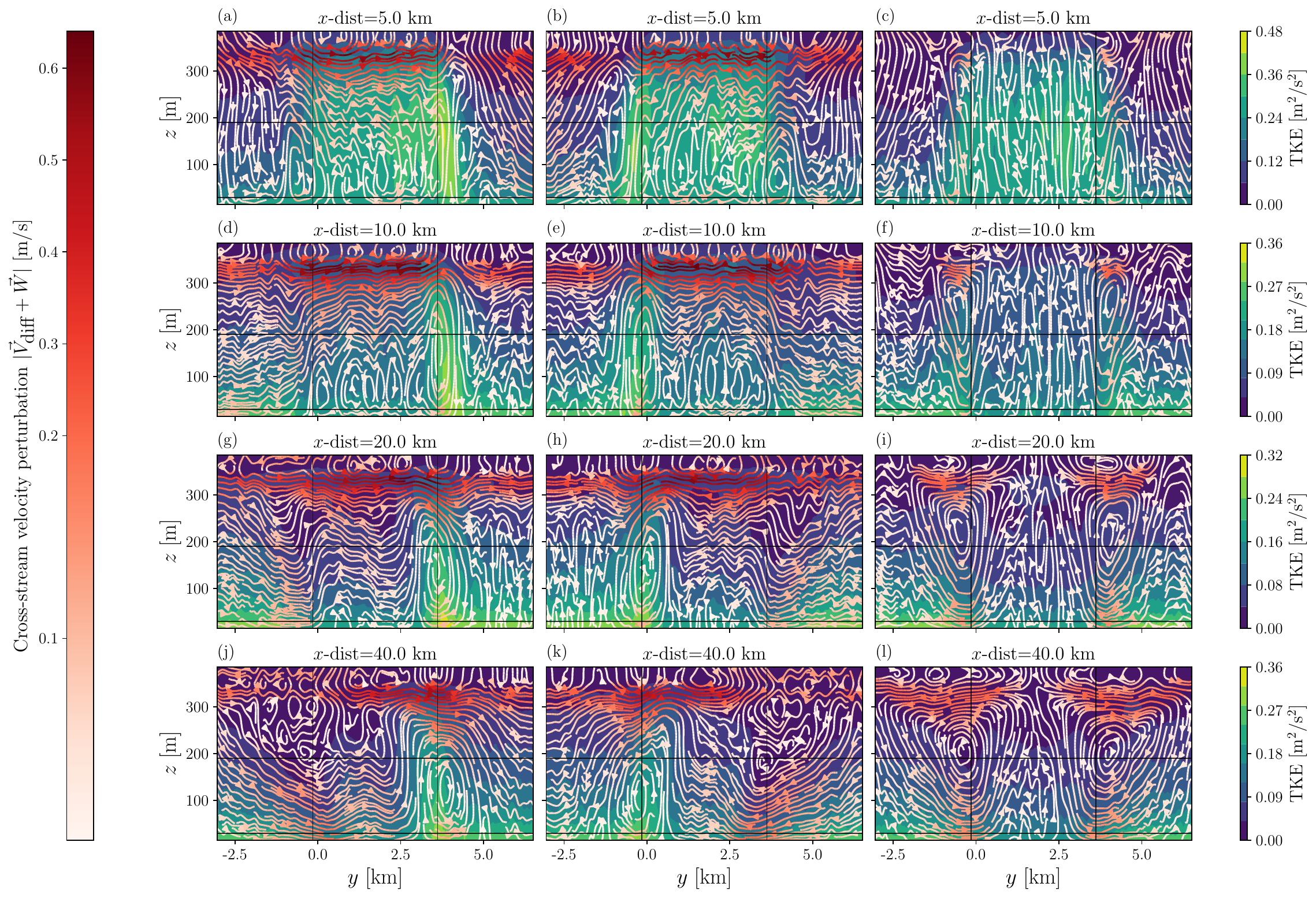}
    \caption{Streamlines of the mean cross-stream velocity perturbation $V_\textrm{diff}$ and $W$ overlaying TKE for several distances downstream of the last turbine of the simulated wind farm for the NH (a,d,g,j), SH (b,e,h,k), and the nE-nC (c,f,i,l) simulations. The streamlines are colour-coded to display the magnitude of the cross-stream velocity perturbation.}
    \label{fig:streamlines}
\end{figure}

In \autoref{fig:cross_planes_u}, we found a signature of cross-stream circulation in the perturbation velocity $V_\textrm{diff}$.
In \autoref{fig:streamlines}, we further investigate this phenomenon by displaying mean flow streamlines for the perturbation velocity in the cross-stream plane (note that $W_\text{in} = 0$, so $W$ is an intrinsic perturbation over the free-stream).

\autoref{fig:streamlines} demonstrates that secondary flows exist at one or both wake edges, depending on the Coriolis effect considered.
In particular, at a distance sufficiently far from the wind farm, a pair of vortices is formed in the nE-nC case at the wake edges, while a single vortex core is found in the NH and SH simulations. These vortex cores resemble the one documented by \citet{Wangsawijaya_2020} and match the direction of rotation, assuming the wake region as a low TKE region and the TKE streak, or the free stream, as a high TKE region. The fact that these structures are not present in the immediate vicinity of the wind farm is probably symptomatic of the large negative gradient of TKE with downstream distance found in the near-wind farm wake.

In our results, all the simulations, but the nE-nC, presented the circulation feature only on the one side of the wake where the TKE streak is expected. We speculated that the tendency of the flow is to form a pair of counter-rotating vortices at both wake edges, as in the nE-nC case. However, the flow convergence along the $y$ direction, caused by the Coriolis effects that we identified as the originating cause of the TKE streak, is likely to interfere with the formation of the vortex pair. Again, the Coriolis effects are responsible for breaking the symmetric behaviour.

We do not investigate these structures further, as, despite influencing the overall farm wake dynamics, they do not seem to impact significantly the presence of the TKE streak, the main topic of the current manuscript. As shown by the fact that the circulations exist in the nE-nC case without promoting TKE streaks.

\bibliographystyle{jfm}
\bibliography{references.bib}

\end{document}